\documentclass[twocolumn,pra,showpacs,floatfix]{revtex4}

\usepackage{graphicx}
\usepackage{dcolumn}
\usepackage{bm}
\usepackage{amsmath}

\begin{document}

\title{Quantal molecular description and universal aspects of the spectra of bosons
and fermions in the lowest Landau level}

\author{Constantine Yannouleas}
\email{Constantine.Yannouleas@physics.gatech.edu}
\author{Uzi Landman}
\email{Uzi.Landman@physics.gatech.edu}

\affiliation{School of Physics, Georgia Institute of Technology,
             Atlanta, Georgia 30332-0430}

\date{29 November 2009}

\begin{abstract}
Through the introduction of a class of trial wave functions portraying combined rotations and
vibrations of molecules formed through particle localization in concentric 
polygonal rings, a correlated basis is constructed that spans the translationally invariant 
part of the lower-Landau-level (LLL) spectra. These trial functions, referred 
to as rovibrational molecular (RVM) functions, generalize our previous work which focused
exclusively on electronic cusp states, describing them as pure vibrationless rotations.
From a computational viewpoint, the RVM correlated basis enables controlled and 
systematic improvements of the original strongly-correlated variational wave function.  
Conceptually, it provides the basis for the development of a quantal molecular description
for the full LLL spectra. This quantal molecular description is universal, being
valid for both bosons and fermions, for both the yrast and excited states of the
LLL spectra, and for both low and high angular momenta. Furthermore, it follows that all
other translationally invariant trial functions (e.g., the Jastrow-Laughlin, compact 
composite-fermion, or Moore-Read functions) are reducible to a description in terms of an 
excited rotating/vibrating quantal molecule.  
\end{abstract}


\maketitle

\section{Introduction}
\label{intro}

\subsection{Motivation}
\label{intmot}
Following the discovery \cite{tsui82} of the fractional
quantum Hall effect (FQHE) in two-dimensional (2D) semiconductor
heterostructures under high magnetic fields ($B$) in the 1980's, the description of
strongly correlated electrons in the lowest Landau level (LLL) developed
into a major branch of theoretical condensed matter physics \cite{laug8399,
hald83,halp84,trug85,jain89,moor91,ston92,oakn95,pala96,jainbook,
yann02,yann03,yann04,yann07,chan05,chan06,jeon04,jeon07}.
Early on, it was realized that the essential many-body physics in the LLL
could be captured through trial wave functions. Prominent examples are
the Jastrow-type Laughlin (JL) \cite{laug8399}, composite fermion (CF) \cite{jain89},
and Moore and Read's (MR) \cite{moor91} Pfaffian functions, representing
quantum-liquid states \cite{laug8399}. In the last ten years, the field of
semiconductor quantum dots \cite{yann07} helped to focus attention on finite
systems with a small number ($N$) of electrons. Theoretical investigations of
such finite systems led to the introduction of ``crystalline''-type LLL trial
functions referred to as rotating electron molecules \cite{yann02,yann07} (REMs).
In particular in their intrinsic frame of reference, the REMs describe electrons 
localized at the apexes of concentric polygonal-ring configurations 
$(n_1,n_2,...,n_r)$, where $\sum_{q=1}^r n_q=N$ and $r$ is the number of concentric 
rings.

More recently, the emerging field of graphene quantum dots \cite{wuns08,roma09}, 
and the burgeoning field of rapidly rotating trapped ultracold neutral gases 
\cite{mott99,coop99,gunn00,pape01,ueda01,popp04,barb06,baks07,vief08,coop08,
geme08,muell08,gunn08} have generated significant interest pertaining to strongly 
correlated states in the lowest Landau level.
Furthermore, it is anticipated that small (and/or mesoscopic) assemblies of 
ultracold bosonic atoms will become technically available in the near future 
\cite{geme08,muell08,gunn08,bour07} and that they will provide an excellent vehicle 
\cite{popp04,barb06,coop08,geme08,muell08,gunn08,bour07} for experimentally reaching 
exotic phases and for testing the rich variety of proposed LLL trial wave functions.

Despite the rich literature and unabated theoretical interest, a unifying physical (as
well as mathematical) description of the full LLL spectra (including both yrast 
\cite{note3} and all excited states), however, is still missing. In this paper,
a universal theory for the LLL spectra of a finite number of particles valid for 
both statistics (i.e., for both bosons and fermions) is introduced. The LLL spectra 
are shown to be associated with {\it fully quantal\/} \cite{note5} and strongly 
correlated ro-vibrational molecular (RVM) states, i.e., with (analytic) trial 
functions describing vibrational excitations relative to the set of the special
yrast states known as cusp states. 

The cusp states exhibit enhanced stability and 
magic angular momenta (see below), and as such they have attracted considerable 
attention. However, the cusp states represent 
only a small fraction of the LLL spectrum. The molecular trial functions 
associated with them are purely rotational (i.e., vibrationless) 
and were introduced for the case of electrons in Ref.\ \cite{yann02} under the 
name rotating electron molecules (REMs). The corresponding purely rotational bosonic 
{\it analytic\/} trial functions for cusp states [called rotating boson molecules 
(RBMs)] are introduced in this paper; see Section \ref{secrbm}.
More importantly, this paper shows that the quantal molecular description can be extended to
all other LLL states (beyond the special cusp states) by introducing (see Section 
\ref{secrvm}) analytic expressions 
for trial functions representing {\it ro-vibrational excitations\/} of both REMs
and RBMs. These ro-vibrational trial functions include the REM or RBM expressions as a 
special case, and they will be referred to in general as RVM trial functions. 

It is remarkable that the numerical results of the present theory were found
in all tested cases to be amenable (if so desired) to an agreement within machine 
precision with exact-diagonalization (EXD) results, including energies, wave 
functions, and overlaps. This numerical behavior points toward a deeper mathematical 
finding, i.e., that the RVM trial functions for both statistics provide a {\it complete\/} 
and {\it correlated\/} basis (see below) 
that spans the translationally invariant (TI) subspace \cite{trug85} of the LLL 
spectrum. An uncorrelated basis, without physical meaning, built out of 
products of elementary symmetric polynomials is also known to span the (bosonic) 
TI subspace \cite{note4}. 

For the sake of clarity, we comment here on the use of the terms ``correlated functions''
and/or ``correlated basis.'' Indeed, the exact many-body eigenstates are customarily called 
correlated when interactions play a dominant role. Consequently, a basis is called 
correlated when its members incorporate/anticipate effects of the strong two-body 
interaction a priori (before the explicit use of the two-body interaction in an exact 
diagonalization). In this respect, Jastrow-type basis wavefunctions (e.g., the 
Feenberg-Clark method of correlated-basis functions \cite{clark59,clark66,muth00} 
and/or the composite-fermion basis \cite{jeon04,jeon07,jeon04.2,note2}) 
are described as correlated, since the Jastrow factors 
incorporate the effect of a strong two-body repulsion in keeping the interacting particles 
apart on the average. Our RVM basis is referred to as correlated since, in addition to 
keeping the interacting particles away from each other, the RVM functions incorporate the 
strong-two-body-repulsion effect of particle localization in concentric polygonal rings and 
formation of Wigner molecules; this localization effect has been repeatedly demonstrated via
EXD calculations in the past decade (see, e.g., the review in Ref.\ \cite{yann07} and 
references therein). In this spirit, we describe the basis of elementary symmetric 
polynomials as ``uncorrelated,'' since the elementary symmetric polynomials do not 
incorporate/anticipate this dominant effect of a strong two-body repulsion, i.e., that of 
keeping the interacting particles apart.   

We are unaware of any other strongly-correlated functions which span the TI subspace. 
Indeed, although the Jastrow-Laughlin function (used for
describing yrast states) is translationally invariant, its quasi-hole
and quasi-electron excitations are not \cite{trug85}. Similarly, the compact
composite-fermion trial functions are translationally invariant \cite{vief08}, 
but the CF excitations which are needed to complete the CF basis are not 
\cite{jeon04,jeon07,jeon04.2,note1}. The shortcoming of the above well known correlated LLL 
theories to satisfy fundamental symmetries of 
the many-body Hamiltonian represents an unsatisfactory state of affairs, and the
present paper provides a remedy to this effect. 
In this context, we note that although the Moore-Read functions \cite{moor91} are also
translationally invariant, they address only certain specific LLL states and they do not 
form a basis spanning the TI subspace. 

\begin{figure}[t]
\centering\includegraphics[width=8.4cm]{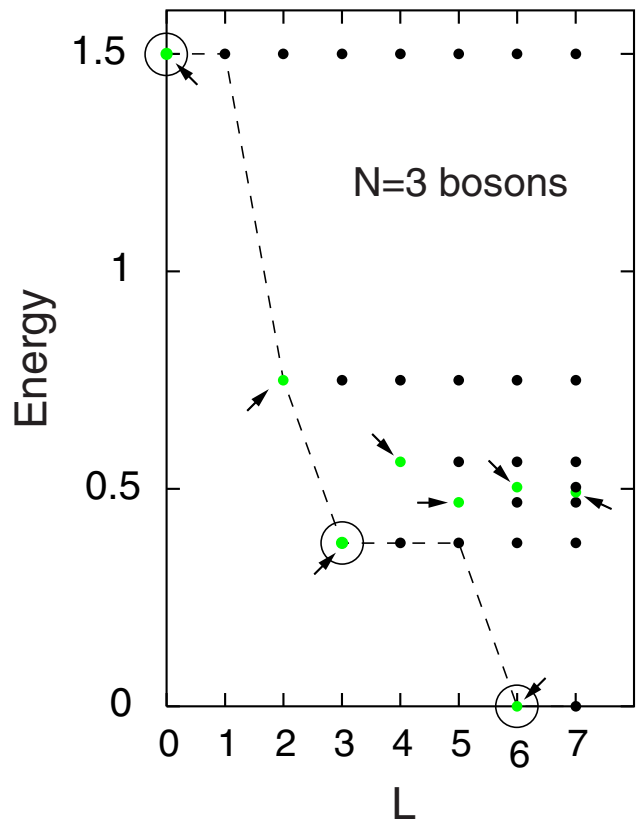}
\caption{(Color online) LLL spectra for $N=3$ scalar bosons calculated using
exact diagonalization. Only the Hamiltonian term containing the two-body repulsive
contact interaction, $g \delta(z_i-z_j)$, [see Eq.\ (\ref{hlll})] was considered in the 
exact diagonalization.
The gray solid dots (marked by arrows; green online) denote the translationally 
invariant states. The dark solid dots are the spurious states (see text). The dashed line 
denotes the yrast band, while the cusp states are marked by a circle. Energies in units of 
$g/(\pi \Lambda^2)$. The number of translationally invariant states is much smaller than the
total number of LLL states.
}
\label{LLLspebos}
\end{figure}

Our introduction of a correlated basis that spans 
the TI subspace is of importance in the following two ways: (1) From a practical 
(and calculational) viewpoint, one can perform controlled and systematic stepwise 
improvements of the original strongly-correlated variational wave function, e.g., the pure 
REM or RBM. (For detailed illustrative examples of the rapid-convergence properties of
the RVM basis, see the Appendix.)
This calculational viewpoint was also the motivation behind the introduction 
of other correlated bases in many-body physics; see, e.g., the treatment of 
quantum liquids and nuclear matter in Refs.\ \cite{clark59,clark66,muth00} and
the composite-fermion correlated basis in Refs.\ \cite{jeon04,jeon07,jeon04.2,note2}.
(2) Conceptually, it guarantees that the properties of the RVM functions, and in
particular the molecular point-group symmetries, are irrevocably incorporated in the
properties of the exact LLL wave functions. Furthermore, it follows that all 
other translationally invariant trial functions (e.g., the JL, compact CF, or Moore-Read 
functions), are reducible to a description in terms of an excited rotating/vibrating 
quantal molecule. Specific examples of the reducibility of the JL and Moore-Read states to 
the molecular description introduced in this paper are provided in Sections 
\ref{4ele} and \ref{5bos}. This is a surprising result, since these Jastrow based trial 
functions are widely described in the previous literature as being liquid-like in an 
essential way.

\subsection{Characteristic properties of the lowest-Landau-level spectra}
\label{intprop}

\begin{figure}[t]
\centering\includegraphics[width=8.4cm]{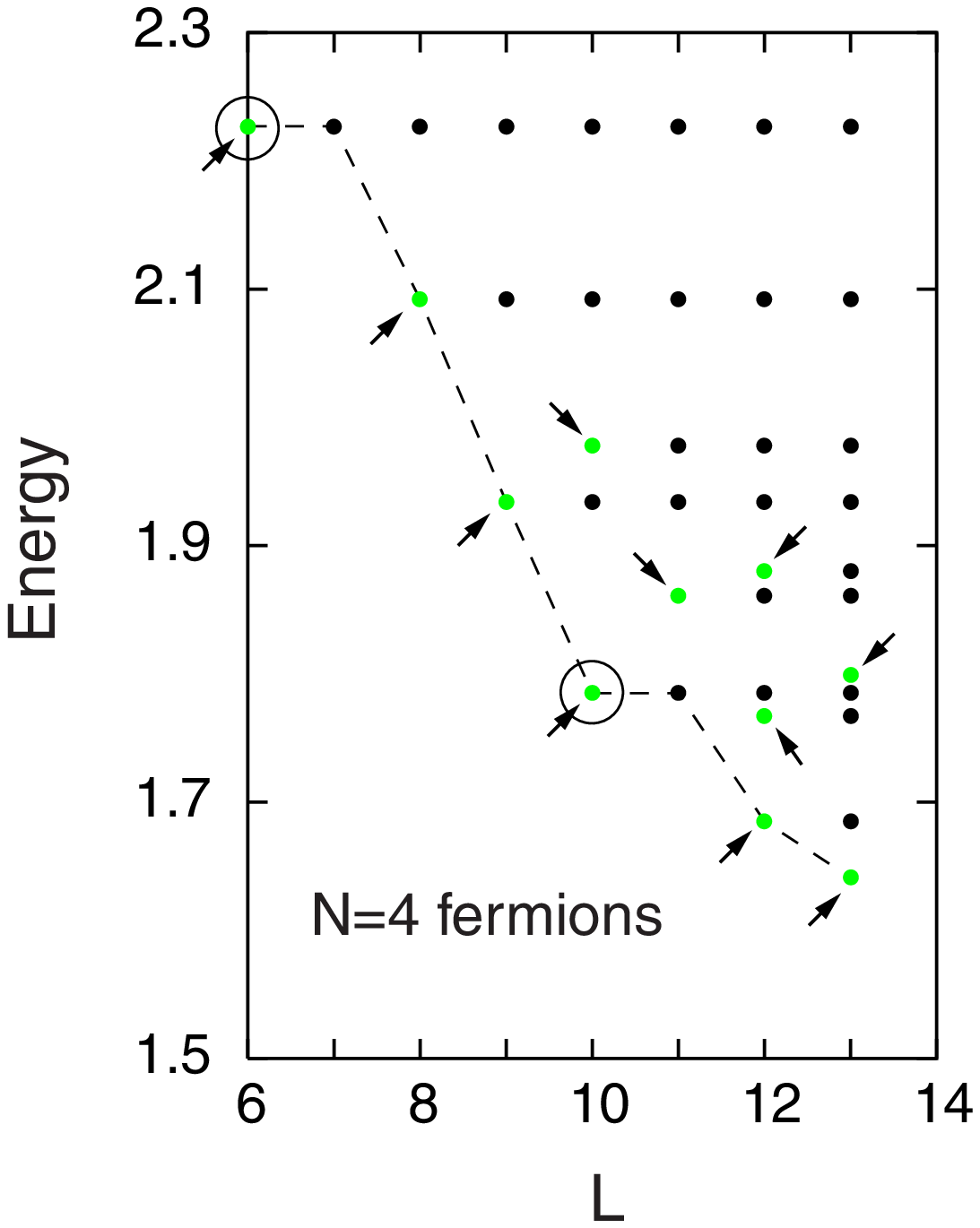}
\caption{(Color online) LLL spectra for $N=4$ spin-polarized electrons calculated using
exact diagonalization. Only the Hamiltonian term containing the two-body Coulomb interaction,
$e^2/(\kappa |z_i-z_j|)$, [see Eq.\ (\ref{hlll})] was considered in the exact 
diagonalization. The gray 
solid dots (marked by arrows; green online) denote the translationally invariant states. 
The dark solid dots are the spurious states (see text). The dashed line denotes the yrast 
band, while the cusp states are marked by a circle. Energies in units of $e^2/(\kappa l_B)$.
The number of translationally invariant states is much smaller than the
total number of LLL states.
}
\label{LLLspeferm}
\end{figure}

For completeness and clarity in the presentation, we briefly provide in this section a 
graphical illustration of some of the main characteristics
of the LLL spectra, calculated via the exact-diagonalization approach. 

First we describe here the special form \cite{jainbook,yann07,vief08} 
of the many-body Hamiltonian
used for calculating the {\it global\/} ground state of a finite number $N$ of electrons at a
given magnetic field $B$. This special form takes advantage of the simplifications at the 
limit of large $B$, i.e., when the relevant Hilbert space can be restricted to the lowest 
Landau level, given that $\hbar \omega_0 << \hbar \omega_c /2$; the frequency $\omega_0$ 
specifies the external harmonic confinement and $\omega_c=eB/(mc)$ is the cyclotron 
frequency. Then the many-body hamiltonian reduces to
\begin{eqnarray}
H^{\text{e,global}}_{\text{LLL}} &=& \nonumber \\
&& \hspace{-2.5cm} N \frac{\hbar \omega_c}{2} 
+ \hbar (\sqrt{\omega_0^2+\omega_c^2/4} -\omega_c/2)L
+ \sum_{i < j}^N \frac{e^2}{\kappa |z_i-z_j|},
\label{hllleg}
\end{eqnarray}
where $L=\sum_{i=1}^N l_i$ is the total angular momentum and $z=x+i y$.

In the case of $N$ rapidly rotating bosons (with $\omega_0$ specifying the external 
confinement of the two-dimensional harmonic trap and $\Omega$ denoting the rotational 
frequency), the corresponding Hamiltonian \cite{gunn00,pape01,baks07,ho01}
(in the limit $\Omega/\omega_0 \rightarrow 1$) is 
written as \cite{note6}
\begin{eqnarray}
H^{\text{b,global}}_{\text{LLL}}&=& \nonumber \\
&& \hspace{-1.5cm} N \hbar \omega_0 + \hbar(\omega_{0}-\Omega) L
+\sum_{i < j}^{N} g \delta(z_i-z_j).
\label{hlllbg}
\end{eqnarray}

Since we will consider many-body energy eigenstates that are eigenstates of the total angular 
momentum as well, it follows that only the interaction terms are nontrivial in both 
Hamiltonians (\ref{hllleg}) and (\ref{hlllbg}). 
As a result, we will henceforth follow the practice of focussing on the
simpler interaction-only LLL Hamiltonian 
\begin{equation}
H_{\text{LLL}}=\sum_{i < j}^{N} v(z_i-z_j),
\label{hlll}
\end{equation}
where $v(z_i-z_j)$ denotes the two-body interaction (Coulomb for electrons
and repulsive contact potential for bosons). The ``ground states'' of the Hamiltonian
in Eq. (\ref{hlll}) coincide with the ``yrast'' band \cite{note3}.

We remind the reader that the EXD method is based on the fact that the full LLL Hilbert space 
at a given total angular momentum $L$ is spanned by the set of all 
possible uncorrelated permanents (for bosons) or Slater determinants (for electrons) made 
out from the Darwin-Fock zero-node single-particle levels (referred to also as orbitals)
\begin{equation}
\psi_{l_i}(z) = \frac{ z^{l_i} } { \sqrt{ \pi {l_i}!} } \exp(-zz^*/2),
\label{lll}
\end{equation}
with $l_i \geq 0$. The position variable $z$ is given in units of
$\Lambda = \sqrt{\hbar/(m\omega_0)}$ in the case of a rotating harmonic trap (with 
lateral confinement frequency $\omega_0$) or $l_B \sqrt{2}$ in the case of an applied 
magnetic field $B$, with $l_B =\sqrt{\hbar/(m\omega_c)}$ being the magnetic length 
and $\omega_c$ the cyclotron frequency \cite{note6}.  
(For details concerning the EXD method, see, e.g., Refs.\ \cite{yann07,baks07}.)
In the following, we use the convention that an uncorrelated state is described
by a single permanent (or Slater determinant) made out from the orbitals
in Eq.\ (\ref{lll}), which are characterized by good single-particle angular momenta $l_i$.

A small part (sufficient for our purposes here) of the EXD LLL spectra (as a function of
$L$) are plotted in Fig.\ \ref{LLLspebos} for $N=3$ scalar bosons and in Fig.\ 
\ref{LLLspeferm} for $N=4$ spin-polarized electrons. As is usually done in the LLL, the 
one-body terms of the Hamiltonian (i.e., confining potential and kinetic-energy) were omitted 
\cite{jainbook,yann07,vief08}, and the exact diagonalization involved only the two-body 
interaction [see Eq.\ (\ref{hlll})].

In Figs.\ \ref{LLLspebos} and \ref{LLLspeferm}, the yrast bands \cite{note3} 
are denoted by a dashed line. Along the
yrast bands there appear special cusp states denoted by a circle. The cusp states are 
important because they exhibit enhanced stability when the one-body terms of the Hamiltonian 
(i.e., external confinement and kinetic energy) are added [see Eqs. (\ref{hllleg}) and 
(\ref{hlllbg})], and thus they determine \cite{note82} the {\it global ground states\/} 
\cite{jainbook,yann07,gunn00,baks07,jain95} as a function of the applied magnetic field $B$ 
or the rotational frequency $\Omega$ of the trap (for the 
correspondence between $B$ and $\Omega$, see Ref.\ \cite{note6}).
In all studied cases \cite{jainbook,yann07,gunn00,barb06,baks07,coop08,jain95} (including both
electrons and bosons up to $N=9$ particles), the total angular momenta of the global
ground states belong to the set of magic angular momenta given by Eq.\ (\ref{mam}) below.
We note that the emergence of these magic angular momenta are a direct signature of the 
molecular nature of the cusp states, a fact that further motivates our present investigations 
concerning the molecular description of the full LLL spectra beyond the electronic
\cite{yann02,yann03,yann07} cusp states. 

In the LLL, all many-body wave functions have the general form\cite{trug85,pape01,vief08}
\begin{equation}
W(z_1,z_2,\ldots,z_N) |0\rangle,
\label{genwf} 
\end{equation}
where $W(z_1,z_2,\ldots,z_N)$ is an homogeneous polynomial of degree $L$ (being antisymmetric
for fermions and symmetric for bosons).

In Eq.\ (\ref{genwf}), the symbol $|0\rangle$ stands for a product of Gaussians [see Eq.\
(\ref{lll})], i.e., 
\begin{equation}
|0\rangle = \exp(-\sum_{i=1}^N z_i z_i^*/2).
\label{prod0}
\end{equation}
To simplify the notation, this common factor will be omitted henceforth in the algebraic 
expressions and manipulations [except in Eqs.\ (\ref{exppsi}) and (\ref{mol_trial_wf})], where
it is repeated for clarity]. Its contribution, however, is necessary when numerical results 
are calculated.

A central property of the LLL spectra is the existence of a translationally invariant 
(TI) subspace \cite{trug85,pape01,vief08} for a given $L$. This subspace is associated with
a special subset of the general wave functions in Eq.\ (\ref{genwf}), i.e., with wave 
functions having translationally invariant polynomials $W(z_1,z_2,\ldots,z_N)$. Specifically,
the TI polynomials obey the relationship 
\begin{equation}
W(z_1+c,z_2+c,\ldots,z_N+c) = W(z_1,z_2,\ldots,z_N),
\label{polti}
\end{equation}
for any arbitrary constant complex number $c$.

The LLL states belonging to the TI subspaces are denoted by gray solid dots 
(marked by an arrow; green online) in Figs.\ 
\ref{LLLspebos} and \ref{LLLspeferm}. The dimension $D^{\text{TI}}(L)$ of the translational 
invariant subspace is much smaller than the dimension $D^{\text{EXD}}(L)$ of the 
exact-diagonalization (EXD) \cite{yann07} space (which is spanned by uncorrelated 
permanents or Slater determinants as discussed above).
The remaining $D^{\text{EXD}}(L) - D^{\text{TI}}(L)$ states are 
{\it spurious\/} center-of-mass excitations, generated by multiplying the TI states with the 
operator $z_c^m$, $m=0,1,2,\ldots$, where 
\begin{equation}
z_c=\frac{1}{N}\sum_{i=1}^N z_i,
\label{zcm}
\end{equation}
is the coordinate of the center of mass \cite{note78}.

The energies of these spurious states coincide with those appearing at all the other 
smaller angular momenta \cite{trug85}. Thus (see TABLES \ref{ene_bos_np3},
\ref{ene_ferm_np4}, and \ref{ene_bos_np5}) 
\begin{equation}
D^{\text{TI}}(L)=D^{\text{EXD}}(L)- D^{\text{EXD}}(L-1).
\label{drel}
\end{equation}

We further note that for $N$ particles (bosons or fermions)
\begin{equation}
D^{\text{TI}}_b(L)=D^{\text{TI}}_f \biglb( L+N(N-1)/2 \bigrb),
\label{dtibf}
\end{equation}
where the subscripts $b$ and $f$ stand for bosons and fermions (electrons), respectively.
$N(N-1)/2$ is the smallest value of the total angular momentum for spin polarized
fermions in the LLL.

\subsection{Plan of the paper}
\label{secplan}

The paper is organized as follows:

The analytic trial functions associated with pure rotations of bosonic molecules (i.e., the
RBMs) are introduced in Section \ref{secrbm}, followed by a description of the purely
rotational electronic molecular functions (i.e., the REMs) in Section \ref{secrem}. 

Properties of the RBMs and REMs are discussed in Section \ref{proprbmrem}. 

Section \ref{secrvm} introduces the general ro-vibrational trial functions (i.e., the
RVMs).

Case studies of the quantal molecular description of the LLL spectra are presented in
Section \ref{secmoldes}. In particular, Section \ref{3bos} discusses the case of $N=3$
LLL scalar bosons, while Section \ref{3ele} discusses the case of $N=3$ spin-polarized LLL 
electrons. The case of $N=4$ LLL electrons is presented in Section \ref{4ele}, along with
an analysis of the Jastrow-Laughlin state (for fractional filling $\nu=1/3$) from the
viewpoint of the present molecular theory. Section \ref{5bos} investigates the case of 
$N=5$ LLL bosons, along with an analysis of the Moore-Read state according to the molecular 
picture.

Section \ref{seccon} offers a summary and discussion.

Finally, the Appendix discusses the rapid-convergence properties of the RVM basis.

We note that, going from $N=3$ to $N=5$ particles, the molecular description requires
consideration of successively larger numbers of isomeric molecular structures as elaborated 
in Section \ref{secmoldes}. In particular, for $N=3$ only one molecular isomer is needed, 
while three different molecular isomers are needed for $N=5$. It is remarkable that these 
isomers are independent of the statistics (bosons or fermions). 

\section{Molecular trial functions}
\label{secmoltrf}

The molecular trial functions introduced in this paper are derived with the help of a 
first-principles methodology of hierarchical successive approximations which 
converge to the exact solution of the many-body Schr\"{o}dinger equation \cite{yann07}. 
Specifically, this methodology is based on the theory of symmetry
breaking at the mean-field level and of subsequent symmetry restoration via
projection techniques \cite{yann07}.
In this Section, we present (and/or review where appropriate) this derivation in some detail.

\subsection{Purely rotational bosonic trial functions (RBMs)}
\label{secrbm}

RBM analytical wave functions in the LLL for $N$ bosons in 
two-dimensional rotating traps can be derived following earlier analogous derivations 
for the case of electrons \cite{yann02}. Our approach consists of two steps:

(I) At the {\it first step\/}, one constructs a permanent (Slater determinant for
fermions) $\Psi^N(z_1,\ldots,z_N)$ out of displaced single-particle states
$u(z_j,Z_j)$, $j=1,\ldots,N$ that represent scalar bosons localized at the positions 
$Z_j$, with (omitting the particle indices) $z=x+iy=$ and $Z=X+iY=R e^{i\phi}$.
\begin{equation}
\Psi^{N} [z] = \text{perm}({\cal M}^N [z]),
\label{perm1}
\end{equation}
with the matrix ${\cal M}^N[z]$ being
\begin{eqnarray}
{\cal M}^N [z] = 
\left[
\begin{array}{ccc}
u_(z_1,Z_1) & \dots & u_(z_N,Z_1) \\
\vdots & \ddots & \vdots \\
u(z_1,Z_N) & \dots & u(z_N,Z_N) \\
\end{array}
\right].
\label{perm2}
\end{eqnarray}
For the permanent of a matrix, we follow here the definition in Ref.\ \cite{wolf1},
that is, the permanent is an analog of a determinant where all the signs in the 
expansion by minors are taken as positive. This definition provides an unnormalized
expression for the permanent. If a normalized expression is needed, one has to
multiply with a normalization constant
\begin{equation}
{\cal N} = \frac{1}{\sqrt{N! p_1! p_2! \ldots p_M!}},
\label{normperm}
\end{equation}
where $\{p_1, p_2, \ldots, p_M\}$ denote the occupations (multiplicities) of the 
orbitals, assuming that there are $M$ distinct orbitals in a given permanent 
($M \leq N$) \cite{note45}. 

In the LLL, one can specifically consider the limit when the confining potential can 
be neglected compared to the effect induced by the gauge field. The localized 
$u(z,Z)$ single-particle states (referred to also as orbitals) are then taken to be 
{\it displaced\/} zero-node Darwin-Fock states with appropriate Peierls 
phases due to the presence of a perpendicular magnetic field [see Eq.\ (1) in Ref.\
\onlinecite{yann02}], or due to the rotation of the trap with angular frequency
$\Omega$. Then, assuming a symmetric gauge, the orbitals can be represented 
\cite{yann02,yann03,mz83} by displaced Gaussian analytic functions, centered at 
different positions $Z_j \equiv X_j+ Y_j$ according to the equilibrium configuration 
of $N$ classical point charges\cite{lozo87,peet9402} arranged at the vertices of 
nested regular polygons (each Gaussian representing a localized particle). Such 
displaced Gaussians are written as
\begin{eqnarray}
u(z&,&Z_j) = (1/\sqrt{\pi}) \nonumber \\
&\times& \exp[-|z-Z_j|^2/2] \exp[-i (xY_j-yX_j)],
\label{gaus}
\end{eqnarray}
where the phase factor is due to the gauge invariance. $z \equiv x+i y$, 
and all lengths are in units of $\Lambda$ in the case of a rotating trap or
$l_B \sqrt{2}$ in the case of an applied magnetic field; see Section \ref{intprop}.

The localized orbital $u(z,Z)$ can be expanded in a series over the complete set of
zero-node single-particle wave functions in Eq.\ (\ref{lll}). One gets [see
Appendix A in Ref.\ \cite{yann06}]
\begin{equation}
u(z,Z)=\sum_{l=0}^{\infty} C_l(Z) \psi_l(z),
\label{uexp}
\end{equation}
with 
\begin{equation}
C_l(Z)=(Z^*)^l \exp(-ZZ^*/2)/\sqrt{l!}
\label{clz}
\end{equation} 
for $Z \neq 0$. Naturally, $C_0(0)=1$ and $C_{l>0}(0)=0$.

For an $N$-particle system, the bosons are situated at the apexes of 
$r$ concentric regular polygons. The ensuing multi-ring structure is
denoted by $(n_1,n_2,...,n_r)$ with $\sum_{q=1}^r n_q=N$.
The position of the $j$-th electron on the $q$-th ring is given by
\begin{equation}
Z_j^q={\widetilde Z}_q \exp[i 2\pi (1-j)/n_q],\;\;  1 \leq j \leq n_q.
\label{zj1}
\end{equation}

The single permanent $\Psi^N [z]$ represents a {\it static\/} boson
molecule. Using Eq.\ (\ref{uexp}), one finds the
following expansion (within a proportionality constant):
\begin{eqnarray}
\Psi^N [z] =&& \sum_{l_1=0,...,l_N=0}^{\infty} 
\frac{ C_{l_1}(Z_1)C_{l_2}(Z_2) \cdot \cdot \cdot C_{l_N}(Z_N) }
{\sqrt{l_1! l_2! \cdot \cdot \cdot l_N!} }  \nonumber \\
&& \times \; P(l_1,l_2,...,l_N) |0\rangle,
\label{exppsi}
\end{eqnarray}
where $P(l_1,l_2,...,l_N) 
\equiv {\text{perm}}[z_1^{l_1},z_2^{l_2}, \ldots, z_N^{l_N}]$; the elements of the 
permanent are the functions $z_i^{l_j}$ , with $z_1^{l_1},z_2^{l_2},\ldots,z_N^{l_N}$ 
being the diagonal elements. The $Z_k$'s (with $1 \leq k \leq N$) in Eq.\ 
(\ref{exppsi}) are the $Z_j^q$'s of Eq.\ (\ref{zj1}), but relabeled.

In Eq.\ (\ref{exppsi}), the common factor $|0\rangle$ represents the product of Gaussians 
defined in Eq.\ (\ref{prod0}). To simplify the notation, this common factor is usually 
omitted. 

(II) {\it Second step:\/}
In the following, we will continue with the details of the complete derivation in the 
simplest case of a single $(0,N)$ ring. Thus we consider the special case
\begin{equation}
Z_j= R e^{2\pi i (1-j)/N},\;\; 1 \leq j \leq N,
\label{posit}
\end{equation}
where $R$ is the radius of the single ring.

The Slater permanent $\Psi^N [z]$ breaks the rotational symmetry and thus
it is not an eigenstate of the total angular momentum $\hbar \hat{L} =
\hbar \sum_{j=1}^N \hat{l}_j$. However, one can restore \cite{yann02,yann07} the
rotational symmetry by applying onto $\Psi^N [z]$ the projection operator
\begin{equation}
{\cal P}_L \equiv \frac{1}{2\pi} \int_0^{2\pi} d\gamma
e^{i\gamma(\hat{L} - L)},
\label{projl}
\end{equation}
where $\hbar L$ are the eigenvalues of the total angular momentum.

When applied onto $\Psi^N [z]$, the projection operator ${\cal P}_L$ acts as
a Kronecker delta: from the unrestricted sum in Eq.\ (\ref{exppsi}) it picks
up only those terms having a given total angular momentum $L$ (henceforth we
drop the constant prefactor $\hbar$ when referring to angular momenta).
As a result the projected wave function $\Phi^N_L ={\cal P}_L \Psi^N$ is
written as (within a proportionality constant)
\begin{eqnarray}
\Phi^N_L[z]=
\sum_{l_1,\dots,l_N}^{l_1+\dots+l_N=L}
\frac{P(l_1,\dots,l_N)}{l_1!\dots l_N!}
e^{i(\phi_1 l_1+\dots+ \phi_N l_N)},\;\;
\label{phi1}
\end{eqnarray}
with $\phi_j=2\pi(j-1)/N$.

We further observe that it is advantageous to rewrite Eq.\ (\ref{phi1}) by
restricting the summation to the ordered arrangements
$l_1 \leq l_2 \leq \ldots \leq l_N$, in which case we get
\begin{eqnarray}
\hspace{-1.0cm} \Phi^N_L[z]&=&
\sum_{0\leq l_1 \leq l_2 \leq \dots \leq l_N }^{l_1+l_2+\dots+l_N=L}
\frac{P(l_1,\dots,l_N)}{l_1!\dots l_N!} \nonumber \\
&& \hspace{-1.5cm} \times
\frac{\mbox{perm}[e^{i\phi_1 l_1}, e^{i\phi_2 l_2},\ldots,e^{i\phi_N l_N}]}
{p_1! p_2! \ldots p_M!}.
\label{phi2}
\end{eqnarray}
The second permanent in Eq.\ (\ref{phi2}) can be shown \cite{mathematica} to
be equal (within a proportionality constant) to a {\it sum} of cosine
terms times a phase factor 
\begin{equation}
e^{i\pi(N-1)L/N},
\label{phf}
\end{equation}
which is independent of the individual $l_j$'s.

The final result for the $(0,N)$ RBM wave function is (within a proportionality
constant):
\begin{widetext}
\begin{equation}
\Phi^{\text{RBM}}_L (0,N)[z] =
\sum_{0 \leq l_1 \leq l_2 \leq \ldots \leq l_N}^{l_1+l_2+\ldots+l_N=L}
C_b(l_1,l_2, \ldots, l_N) 
{\text{perm}} [z_1^{l_1}, z_2^{l_2}, \ldots, z_N^{l_N}], 
\label{rbm0n}
\end{equation}
\end{widetext}
where the coefficients are given by different expressions for even or odd number
of bosons $N$.

(i) For even $N$, one has

\begin{widetext}
\begin{eqnarray}
C_b(l_1,l_2,\ldots,l_N) = \left(\prod_{i=1}^N l_i! \right)^{-1} 
\left(\prod_{k=1}^M p_k! \right)^{-1} && \nonumber \\
&& \hspace{-7.5cm} \times 
\sum_\sigma  \cos \left\{ 
[ (N-1)l_{\sigma_1}+(N-3)l_{\sigma_2} + \ldots + l_{\sigma_{(N/2)}}
-l_{\sigma_{(N/2+1)}} - \ldots - (N-3)l_{\sigma_{(N-1)}} -(N-1)l_{\sigma_N} ]
\frac{\pi}{N} 
\right\},
\label{rbm0neven}
\end{eqnarray}
\end{widetext}
where the summation $\sum_\sigma$ runs over the permutations of the set of
$N$ indices $\{1,2, \ldots, N\}$.

(ii) With the notation $K=N-1$, the corresponding coefficients for odd $N$ 
are:

\begin{widetext}
\begin{eqnarray}
C_b(l_1,l_2,\ldots,l_N) = \left(\prod_{i=1}^N l_i! \right)^{-1}
\left(\prod_{k=1}^M p_k! \right)^{-1} && \nonumber \\
&& \hspace{-7.5cm} \times
\sum_{ \sigma \{ K \} }  \cos \left\{
[ K l_{\sigma_1}+(K-2)l_{\sigma_2} + \ldots + 2 l_{ \sigma_{(K/2)} }
-2 l_{ \sigma_{(K/2+1)} } - \ldots - (K-2)l_{\sigma_{ (K-1) } } 
-K l_{ \sigma_{ K }} ]
\frac{\pi}{N}
\right\},
\label{rbm0nodd}
\end{eqnarray}
\end{widetext}
where $\sum_{ \sigma \{ K \} }$ runs over all permutations of $N-1$ 
indices selected out from the set $\{1,2, \ldots, N\}$ (of $N$ indices).

In both Eqs.\ (\ref{rbm0neven}) and (\ref{rbm0nodd}), the index $M$ 
(with $1 \leq M \leq N$ denotes the number of different single-particle angular
momenta $l_j$'s ($j=1,2,\ldots,M$) in the ordered list $\{ l_1,l_2,\ldots,l_N \}$ and
the $p_k$'s are the multiplicities of each one of the different $l_j$'s
[occupations of the corresponding single-particle orbitals $\psi_{l_j}(z)$].
For example, for $N=4$ and $\{l_1=2,l_2=2,l_3=2,l_4=5\}$, one has $M=2$ and
$p_1=3$, $p_2=1$; for $\{ l_1=0,l_2=0,l_3=0,l_4=0 \}$, one has $M=1$ and $p_1=4$;
for $\{ l_1=1,l_2=2,l_3=3,l_4=9 \}$, one has $M=4$ and $p_1=p_2=p_3=p_4=1$.

We further note that for both Eqs.\ (\ref{rbm0neven}) and (\ref{rbm0nodd}) the total 
number of distinct cosine terms is $N!/2$ \cite{wolf2}, with the division
by 2 following from the symmetry properties of cosine, i.e., from
$\cos(-x)=\cos(x)$. 

The RBM expresion for an $(n_1,n_2)$ two-ring configuration (with $N=n_1+n_2$) can be
derived following similar steps as in the derivation of the expressions for the 
multi-ring REMs \cite{yann02}. If $L_1+L_2=L$, the final two-ring RBM expression is 
\begin{widetext}
\begin{eqnarray}
\Phi^{\text{RBM}}_{L} (n_1,n_2)[z] &=& \nonumber \\
&& \hspace{-3.2cm}
\sum_{0 \leq l_1 \leq l_2 \leq \ldots \leq l_{n_1}}^{l_1+l_2+\ldots+l_{n_1}=L_1}
\sum_{0 \leq l_{n_1+1} \leq l_{n_2+2} 
\leq \ldots \leq l_N}^{l_{n_1+1}+l_{n_1+2}+ \ldots +l_{N}=L_2}
C_b(l_1,l_2, \ldots, l_{n_1}) C_b(l_{n_1+1},l_{n_1+2}, \ldots, l_{N}) 
{\text{perm}} [z_1^{l_1}, z_2^{l_2}, \ldots, z_N^{l_N}], \nonumber \\
\label{rbmn1n2}
\end{eqnarray}
\end{widetext}
where $C_b(l_1,l_2, \ldots, l_{n_1})$ and $C_b(l_{n_1+1},l_{n_1+2}, \ldots, l_{N})$ 
are calculated by applying the single-ring expressions of Eqs.\ (\ref{rbm0neven}) and 
(\ref{rbm0nodd}).

Generalizations of expression (\ref{rbmn1n2}) to structures with a larger
number $r$ of concentric rings involve for {\it each\/} $q$-th ring ($ 1 \leq q \leq r$):
(I) Consideration of a separate factor 
$ C_b(l_{n_{q-1}+1},l_{n_{q-1}+2}, \ldots, l_{ n_{q-1}+n_{q} })$; 
(II) A restriction on the summation of the associated $n_q$ angular momenta,
i.e., $l_{n_{q-1}+1}+l_{n_{q-1}+2}+ \ldots +l_{n_{q-1}+n_{q}}=L_q$, with 
$\sum_{q=1}^r L_q = L$.

\subsection{Purely rotational fermionic trial functions (REMs)}
\label{secrem}

The REM expresion for any $(n_1,n_2,\ldots,n_r)$ multi-ring configuration (with 
$N=\sum_{q=1}^r n_q$) was derived earlier in Refs.\ \cite{yann02,yann03} following 
similar steps as those in Section \ref{secrbm}. A determinant needs to be used, however,
instead of a permanent, to conform with the antisymmetrization properties of the electronic 
(fermionic) many-body wave function. If $L_1+L_2=L$, the final two-ring REM expression is
\begin{widetext}
\begin{eqnarray}
\Phi^{\text{REM}}_{L} (n_1,n_2)[z] &=& \nonumber \\
&& \hspace{-3.2cm}
\sum_{0 \leq l_1 < l_2 < \ldots < l_{n_1} < l_{n_1+1} < \ldots < l_N}
^{l_1+l_2+\ldots+l_{n_1}=L_1, l_{n_1+1}+l_{n_1+2}+ \ldots +l_{N}=L_2}
C_f(l_1,l_2, \ldots, l_{n_1}) C_f(l_{n_1+1},l_{n_1+2}, \ldots, l_{N})
{\text{det}} [z_1^{l_1}, z_2^{l_2}, \ldots, z_N^{l_N}], \nonumber \\
\label{remn1n2}
\end{eqnarray}
\end{widetext}
where the fermionic coefficients $C_f(l_1,l_2, \ldots, l_{n_1})$ and 
$C_f(l_{n_1+1},l_{n_1+2}, \ldots, l_{N})$ are calculated by applying to each one of them 
the single-ring [$(0,N)$] expression
\begin{eqnarray}
C_f(l_1,l_2, \ldots, l_{N}) &=& \nonumber \\
&& \hspace{-3.2cm} \left( \prod_{i=1}^N l_i! \right)^{-1} 
\left( \prod_{1 \leq i < j \leq N} 
\sin \left[\frac{\pi}{N}(l_i-l_j)\right] \right).  
\label{cf}
\end{eqnarray} 

It is straighforward to generalize the two-ring REM expression in Eq.\ (\ref{remn1n2}) 
to more complicated or simpler [i.e., $(0,N)$ and $(1,N-1)$] configurations by 
(I) considering a separate factor 
$ C_f(l_{n_{q-1}+1},l_{n_{q-1}+2}, \ldots, l_{ n_{q-1}+n_{q} })$ for each $q$th ring;
(II) restricting the summation of the associated $n_q$ angular momenta,
i.e., $l_{n_{q-1}+1}+l_{n_{q-1}+2}+ \ldots +l_{n_{q-1}+n_{q}}=L_q$, with
$\sum_{q=1}^r L_q =L$. 

Apart from the permanent being replaced by a determinant,
we note two other main differences between the REM and RBM expressions. That is,
for electrons (1) $M=N$ in all instances (single occupancy) and (2) a {\it product of 
sine\/} terms in the coefficients $C_f(\ldots)$ replaces the {\it sum of cosine\/} terms in 
the coefficients $C_b(\ldots)$. 

\subsection{Properties of RBMs and REMs}
\label{proprbmrem}
The analytic expressions for $\Phi^{\text{RXM}}_{\cal L}(n_1,n_2,\ldots,n_r)[z]$
(the index RXM standing for either RBM or REM)
describe pure molecular rotations associated with magic angular momenta
\begin{equation}
{\cal L}={\cal L}_0+\sum_{q=1}^r n_q k_q, 
\label{mam}
\end{equation}
with $k_q$, $q=1,\ldots,r$ being nonnegative integers; ${\cal L}_0=N(N-1)/2$ for 
electrons and ${\cal L}_0=0$ for bosons.

A central property of these trial functions is that identically
\begin{equation}
\Phi^{\text{RXM}}_{\cal L}(n_1,n_2,\ldots,n_r)[z]=0
\label{eqzero}
\end{equation}
for both bosons and electrons when 
\begin{equation}
{\cal L} \neq {\cal L}_0 + \sum_{q=1}^r n_q k_q
\label{neqtomagic}
\end{equation}
This selection rule follows directly from the point group symmetries of the 
$(n_1,n_2,\ldots,n_r)$ multi-ring polygonal configurations. Indeed under condition
(\ref{neqtomagic}) the $C_b(\ldots)$ and $C_f(\ldots)$ coefficients are identically
zero, as can be easily checked using MATHEMATICA \cite{mathematica}. In other words,
purely rotational states are allowed only for certain angular momenta that do not 
conflict with the intrinsic molecular point-group symmetries.

The yrast states corresponding to magic ${\cal L}$'s [Eq.\ (\ref{mam})] are associated with 
the special cusp states described previously in Figs.\ \ref{LLLspebos} and \ref{LLLspeferm} 
and in Section \ref{intprop}. Furthermore, the enhanced stability associated with the cusp 
states (see Section \ref{intprop}) is obviously due to the selection rule described by Eqs.\ 
(\ref{eqzero}) and (\ref{neqtomagic}).

An important property of the REM and RBM trial functions is their translational invariance
(in the sense described in Section \ref{intprop}).

\begin{table*}[t] 
\caption{\label{ene_bos_np3}%
LLL spectra of three spinless bosons interacting via a repulsive contact 
interaction $g \delta(z_i-z_j)$. 2nd column: Dimensions of the EXD and 
the nonspurious TI (in parenthesis) spaces (the EXD space is spanned by 
uncorrelated permanents of Darwin-Fock orbitals). 
4th to 6th columns: Matrix elements [in units of $g/(\pi \Lambda^2)$, 
$\Lambda=\sqrt{\hbar/(m\omega_0)}$] of the contact interaction between the 
correlated RVM states $\{k,m \}$ [see Eq.\ (\ref{wfbos3})]. The total angular 
momentum $L=3k+2m$. Last three columns: Energy eigenvalues from the RVM 
diagonalization of the associated matrix of dimension $D^{\text{TI}}(L)$. There 
is no nonspurious state with $L=1$. The full EXD spectrum at a given $L$ is 
constructed by including, in addition to the listed TI eigenvalues 
[$D^{\text{TI}}(L)$ in number], all the energies associated with angular 
momenta smaller than $L$. An integer in square brackets indicates the energy 
ordering in the full EXD spectrum (including both spurious and TI states).
Seven decimal digits are displayed, but the energy eigenvalues from the RVM 
diagonalization agree with the corresponding EXD$^{\text{TI}}$ ones within machine 
precision.}
\begin{ruledtabular}
\begin{tabular}{rllllllll}
$L$ & $D^{\text{EXD}}(D^{\text{TI}})$ & $\{k,m\}$ & 
\multicolumn{3}{l}{Matrix elements} & 
\multicolumn{3}{l}{Energy eigenvalues (RVM diag. or EXD$^{\text{TI}}$)} \\ \hline
0 & 1(1)  & \{0,0\} & 1.5000000  & 
  ~~~~      &  ~~~~ & 1.5000000[1] & ~~~~ & ~~~~\\
2 & 2(1) & \{0,1\} & 0.7500000  &
  ~~~~      & ~~~~ & 0.7500000[1] & ~~~~ & ~~~~\\
3 & 3(1) & \{1,0\} & 0.3750000  &
  ~~~~      & ~~~~ & 0.3750000[1] & ~~~~ & ~~~~\\
4 & 4(1) & \{0,2\} & 0.5625000  &
  ~~~~      & ~~~~ & 0.5625000[2] & ~~~~ & ~~~~\\
5 & 5(1) & \{1,1\} & 0.4687500  & 
  ~~~~      & ~~~~ & 0.4687500[2] & ~~~~ & ~~~~\\
6 & 7(2) & \{2,0\} & 0.0468750  & 
 0.1482318  &  ~~~~ &  ~~~~       & ~~~~ & ~~~~\\
~~~   & ~ & \{0,3\} & 0.1482318  & 0.4687500   & ~~~~ & 0.0000000[1] & 
0.5156250[4] & ~~~~\\
7 & 8(1) & \{1,2\} & 0.4921875  & 
  ~~~~      & ~~~~ & 0.4921875[4] & ~~~~ & ~~~~\\
8 & 10(2) & \{2,1\} & 0.0937500  & 
0.1960922   &~~~~ &  ~~~~         & ~~~~ & ~~~~\\
~~~   & ~ & \{0,4\} & 0.1960922  & 0.4101562  & ~~~~ & 0.0000000 & 
0.5039062[6] & ~~~~\\
12 & 19(3) & \{4,0\} & 7.3242187$\times 10^{-4}$ & 1.0863572$\times 10^{-2}$ & 
1.5742811$\times 10^{-2}$ & ~~~~~~  & ~~~~~~  & ~~~~~ \\
~~~ & ~~~ & \{2,3\} & 1.0863572$\times 10^{-2}$ & 0.1611328 & 
0.2335036 & ~~~~~~ & ~~~~~  & ~~~~~ \\
~~~ & ~~~ & \{0,6\} & 1.5742811$\times 10^{-2}$ & 0.2335036 & 
0.3383789 & 0.0000000  & 0.0000000  & 0.5002441[13] \\
\end{tabular}
\end{ruledtabular}
\end{table*}

\subsection{General ro-vibrational trial functions (RVMs)}
\label{secrvm}

The RVM functions that account for the general excitations of the rotating molecules 
have the form (within a normalization constant):
\begin{equation}
\Phi^{\text{RXM}}_{\cal L}(n_1,n_2,\ldots,n_r) Q_\lambda^m |0\rangle,
\label{mol_trial_wf}
\end{equation}
The index RXM stands for either REM, i.e., a rotating electron molecule, or RBM, i.e., 
a rotating boson molecule. The purely rotational functions 
$\Phi^{\text{RXM}}_{\cal L}(n_1,n_2,\ldots,n_r)$ have been described in detail in 
earlier sections. The product in Eq.\ (\ref{mol_trial_wf}) combines rotations 
with vibrational excitations, the latter being 
denoted by $Q_\lambda^m$, with $\lambda$ being an angular momentum; the 
superscript denotes raising to a power $m$. Both $\Phi^{\text{RXM}}_{\cal L}$ 
and $Q_\lambda^m$ are homogeneous polynomials of the complex particle coordinates 
$z_1,z_2,\ldots,z_N$, of order ${\cal L}$ and $\lambda m$, respectively.
The total angular momentum $L={\cal L}+\lambda m$. $Q_\lambda^m$ is always 
symmetric in these variables; $\Phi^{\text{RXM}}_{\cal L}$ is antisymmetric 
(symmetric) for fermions (bosons). $|0\rangle$ is the product of Gaussians defined in
Eq.\ (\ref{prod0}); this product of Gaussians is usually omitted.

The vibrational excitations $Q_\lambda$ are given by the same expression for
both bosons and electrons, namely, by the symmetric polynomials:
\begin{equation}
Q_\lambda = \sum_{i=1}^N (z_i-z_c)^\lambda,
\label{ql} 
\end{equation}
where $z_c$ is the coordinate of the center of mass defined in Eq.\ (\ref{zcm})  and 
$\lambda>1$ is an integer positive number. Vibrational excitations of a similar form, i.e., 
\begin{equation}
\tilde{Q}_\lambda=\sum_{i=1}^N z_i^\lambda
\label{qlam2}
\end{equation}
(and certain other variants), 
have been used earlier to approximate {\it part\/} of the LLL spectra. Such 
earlier endeavors provided valuable insights, but overall they remained 
inconclusive; for electrons over the maximum density droplet [with magic 
${\cal L}={\cal L}_0$], see Refs.\ \cite{ston92} and \cite{oakn95}; for electrons 
over the $\nu=1/3$ ($\nu={\cal L}_0/{\cal L}$) Jastrow-Laughlin trial function 
[with magic ${\cal L}=3 {\cal L}_0$], see Ref.\ \cite{pala96}; and for bosons in 
the range $0 \leq L \leq N$, see Refs.\ \cite{mott99,ueda01,pape01}. 

The advantage of $Q_\lambda$ \cite{note11} (compared to $\tilde{Q}_\lambda$) is that it 
is translationally invariant (TI) \cite{trug85,pape01}, a property shared with both  
$\Phi^{\text{RBM}}_{\cal L}$ and $\Phi^{\text{REM}}_{\cal L}$.
In the following, we will discuss illustrative cases, which will show that
the RVM functions of Eq.\ (\ref{mol_trial_wf}) provide a correlated
basis (RVM basis) that spans the TI subspace \cite{trug85,pape01,vief08} of 
{\it nonspurious\/} states in the LLL spectra. 

\section{Molecular description of LLL spectra}
\label{secmoldes}

\subsection{Three spinless bosons}
\label{3bos}

Only the $(0,3)$ molecular 
configuration and the dipolar $\lambda=2$ vibrations are at play
(as checked numerically), i.e.,
the full TI spectra at any $L$ are spanned by the wave functions 
\begin{equation}
\Phi^{\text{RBM}}_{3k}(0,3) Q_2^m \Rightarrow \{k,m\},
\label{wfbos3}
\end{equation} 
with $k,m=0,1,2,\ldots$, and $L=3k+2m$; these states 
are always orthogonal. 

Following Eq.\ (\ref{rbm0n}), a simplified analytic expression for the (0,3) RBM can be
derived, i.e., 
\begin{equation}
\Phi^{\text{RBM}}_{\cal L}(0,3) = 
\sum_{0 \leq l_1 \leq l_2 \leq l_3}^{l_1+l_2+l_3={\cal L}}
C_b(l_1,l_2,l_3) \;{\text{Perm}} [z_1^{l_1}, z_2^{l_2}, z_3^{l_3}],
\label{rbm1n3}
\end{equation}
where the coefficients $C_b(\ldots)$ are given by:
\begin{eqnarray}
C_b(l_1,l_2,l_3) &=& \left(\prod_{i=1}^3 l_i! \right)^{-1} 
\left(\prod_{k=1}^M p_k! \right)^{-1}
\nonumber \\
&\times& \left( \sum_{1 \leq i < j \leq 3} 
\cos \left[\frac{2\pi(l_i-l_j)}{3} \right] \right),
\label{rbm2n3}
\end{eqnarray} 
where $1 \leq M \leq 3$ denotes the number of different single-particle angular momenta
in the triad $(l_1,l_2,l_3)$ and the $p_k$'s are the multiplicities of each one of these 
different angular momenta.

TABLE \ref{ene_bos_np3} provides the systematics of the molecular
description for the beginning ($0 \leq L \leq 12$) of the LLL spectrum. 
There are several cases when the TI subspace has dimension one and the exact solution 
$\Phi^{\text{exact}}$ coincides with a single
$\{k,m\}$ state. For $L=0$ the exact solution coincides with 
$\Phi^{\text{RBM}}_0=1$ ($Q_\lambda^0=1$); this is the only case when an LLL state 
has a Gross-Pitaevskii form, i.e., it is a single (normalized) permanent [see Eq.\ 
(\ref{mol_trial_wf})] given by $|0\rangle$ as defined in Eq.\ (\ref{prod0}).

For $L=2$, we found $\Phi^{\text{exact}}_{[1]} \propto Q_2$, with the index $[i]$ 
indicating the energy ordering in the full EXD spectrum (including both spurious and 
TI states). Since [see Eq.\ (\ref{ql})]
\begin{equation}
Q_2 \propto(z_1-z_c)(z_2-z_c)+(z_1-z_c)(z_3-z_c)+(z_2-z_c)(z_3-z_c),
\label{q2n3}
\end{equation}
this result agrees with the findings of Refs.\ \cite{smit00,pape01} concerning 
ground states of bosons in the range $0 \leq L \leq N$. 

For $L=3$, one finds $\Phi^{\text{exact}}_{[1]} \propto \Phi^{\text{RBM}}_3$. Since 
[see Eq.\ (\ref{rbm1n3})] 
\begin{equation}
\Phi^{\text{RBM}}_3 \propto(z_1-z_c)(z_2-z_c)(z_3-z_c),
\label{rbml3n3}
\end{equation}
this result agrees again with the findings of Refs.\ \cite{smit00,pape01}. 

For $L=5$, the single nonspurious state is an excited one, 
$\Phi^{\text{exact}}_{[2]} \propto \Phi^{\text{RBM}}_3 Q_2$.

For $L=6$ ($\nu=1/2$), the ground-state is found to be 
\begin{eqnarray}
\Phi^{\text{exact}}_{[1]} & \propto & 
-\frac{160}{9} \Phi^{\text{RBM}}_6  + \frac{1}{4} Q_2^3 \nonumber \\
& = &(z_1-z_2)^2(z_1-z_3)^2(z_2-z_3)^2, 
\label{rbml6n3}
\end{eqnarray}
i.e., the bosonic Jastrow-Laughlin function for $\nu=1/2$ is equivalent to an RBM state that 
incorporates vibrational correlations. 

For $L \geq N(N-1)$ (i.e., $\nu \leq 1/2$), the EXD yrast 
energies equal zero, and with increasing $L$ the degeneracy of the zero-energy 
states for a given $L$ increases. It is important that this nontrivial behavior 
is reproduced faithfully by the present method (see TABLE \ref{ene_bos_np3}).

\subsection{Three electrons}
\label{3ele}

Although unrecognized, the solution of the problem of three spin-polarized electrons in the 
LLL using molecular trial functions was presented by Laughlin in Ref.\ \cite{laug83.2}. 
Indeed, the main result of Ref.\ \cite{laug83.2} [see Eq.\ (18) therein] were the following 
wave functions (we display the polynomial part only)
\begin{eqnarray}
|k,m\rangle &\propto&~~~~~~~~~ \nonumber \\
&& \hspace{-1.5cm} \left[ \frac{(z_a+iz_b)^{3k}-(z_a-iz_b)^{3k}}{2i} \right]
(z_a^2+z_b^2)^m, 
\label{laug3}
\end{eqnarray}
where the three-particle Jacobi coordinates are
\begin{equation}
z_c= \frac{z_1+z_2+z_3}{3},
\label{jzc} 
\end{equation}\begin{equation}
z_a=\left( \frac{2}{3} \right)^{1/2} \left[ \frac{z_1+z_2}{2}-z_3 \right],
\label{jza} 
\end{equation}
\begin{equation}
z_b= \frac{1}{\sqrt{2}}(z_1-z_2).
\label{jzb} 
\end{equation}
Expression (\ref{laug3}) is precisely of the form 
$\Phi^{\text{REM}}_{3k} Q_2^m$, as can be checked after transforming back to 
Cartesian coordinates $z_1$, $z_2$, and $z_3$. Thus the wave functions $|k,m\rangle$
of Ref.\ \cite{laug83.2} describe both pure molecular rotations, as well as vibrational 
excitations, and they cover the translationally invariant LLL subspace. We note that the 
pairs of indices $\{k,m\}$ are universal and independent of the statistics, i.e., the same 
for both bosons [Eq.\ (\ref{wfbos3})] and electrons [Eq.\ (\ref{laug3})], as can be 
explicitly seen through a comparison of TABLE \ref{ene_bos_np3} here and TABLE I in Ref.\ 
\cite{laug83.2}.  

We further note that Laughlin did not present molecular trial functions for electrons with 
$N >3$, or for bosons for any $N$. This is done in the present paper. 

\subsection{Four electrons}
\label{4ele}

\begin{table*}[t] 
\caption{\label{ene_ferm_np4}%
LLL spectra of four spin-polarized electrons interacting via the Coulomb
repulsion $e^2/(\kappa |z_i-z_j|)$ . Second column: Dimensions of the full EXD
and the nonspurious TI (in parenthesis) spaces (the EXD space is spanned by
uncorrelated determinants of Darwin-Fock orbitals). Last three columns: Energy
eigenvalues [in units of $e^2/(\kappa l_B)$] from the diagonalization of the
Coulomb interaction in the TI subspace spanned by the trial functions
$\Phi^{\text{REM}}_{6+4k}(0,4) Q_\lambda^m$ and
$\Phi^{\text{REM}}_{6+3k}(1,3) Q_\lambda^m$ (RVM digonalization).
Third to sixth columns: the molecular configurations $(n_1,n_2)$ and the
quantum numbers $k$, $\lambda$ and $m$ are indicated within brackets.
There is no nonspurious state with $L=7$. The full EXD spectrum at a given $L$
is constructed by including, in addition to the listed TI energy
eigenvalues [$D^{\text{TI}}(L)$ in number], all the energies associated with
angular momenta smaller than $L$. An integer in square brackets indicates the
energy ordering in the full EXD spectrum (including both spurious and TI
states), with [1] denoting an yrast state. Eight decimal digits are displayed,
but the energy eigenvalues from the RVM diagonalization agree with the
corresponding EXD$^{\text{TI}}$ ones within machine precision.}
\begin{ruledtabular}
\begin{tabular}{llllll}
$L$ & $D^{\text{EXD}}$($D^{\text{TI}}$)& $[(n_1,n_2)\{ k, \lambda, m \}]$ &
\multicolumn{3}{l}{Energy eigenvalues (RVM diag. or EXD$^{\text{TI}}$)} \\
\hline
6 & 1(1) & [(0,4)\{0,$\lambda$,0\}] &
2.22725097[1]  &   ~~~~      & ~~~~~ \\
8 & 2(1) & [(0,4)\{0,2,1\}] &
2.09240211[1] &   ~~~~      & ~~~~~~ \\
9 & 3(1) & [(1,3)\{1,$\lambda$,0\}] &
 1.93480798[1] &   ~~~~      & ~~~~ \\
10 & 5(2) & [(0,4)\{1,$\lambda$,0\}] [(0,4)\{0,2,2\}] &
 1.78508849[1]  &  1.97809256[3] & ~~~~ \\
11 & 6(1) & [(1,3)\{1,2,1\}] & 1.86157215[2]  &
  ~~~~      & ~~~~~ \\
12 & 9(3) &  [(0,4)\{1,2,1\}] [(0,4)\{0,2,3\}] [(1,3)\{2,$\lambda$,0\}] &
 1.68518201[1] & 1.76757420[2]  & 1.88068652[5] \\
13 & 11(2) & [(1,3)\{1,2,2\}] [(0,4)\{1,3,1\}]  & 1.64156849[1]  &
 1.79962234[5] & ~~~~ \\
14 & 15(4) & [(0,4)\{2,$\lambda$,0\}] [(0,4)\{1,2,2\}] [(0,4)\{0,2,4\}] &
 1.50065835[1]  & 1.63572496[2]   &  1.72910626[5]  \\
~~~ & ~~~~ & [(1,3)\{2,2,1\}] &
 1.79894008[8] & ~~~~~   & ~~~~~~ \\
15 & 18(3) & [(1,3)\{3,$\lambda$,0\}] [(1,3)\{2,3,1\}] [(1,3)\{1,3,2\}]  &
 1.52704695[2] & 1.62342533[3] & 1.74810279[8] \\
18 & 34(7) & [(0,4)\{3,$\lambda$,0\}] [(0,4)\{2,2,2\}] [(0,4)\{1,2,4\}] &
 1.30572905[1]  &  1.41507954[2]   &  1.43427543[4] \\
~~~~ & ~~~~ & [(0,4)\{0,2,6\}] [(1,3)\{4,$\lambda$,0\}] [(1,3)\{2,2,3\}]  &
 1.50366728[8]  &   1.56527615[11]   & 1.63564655[15] \\
~~~~ & ~~~~ & [(1,3)\{3,3,1\}]  &
 1.68994048[20]  &   ~~~~      & ~~~~~ \\
\end{tabular}
\end{ruledtabular}
\end{table*}

For $N=4$ spin-polarized electrons, one needs to
consider two distinct molecular configurations, i.e., $(0,4)$ and $(1,3)$.
Vibrations with $\lambda \geq 2$ must also be considered. In this case the 
RVM states are not always orthogonal, and the Gram-Schmidt orthogonalization is 
implemented.

\begin{table}[b] 
\caption{\label{exp_coeff}%
$N=4$ LLL electrons with $L=18$: Expansion coefficients in the RVM basis (labelled by 
the $|i\rangle$'s) for the three lowest-in-energy EXD$^{\text{TI}}$ states (labelled
[1], [2], [4]; see TABLE \ref{ene_ferm_np4}). The 4th column gives the RVM expansion 
coefficients of the corresponding Jastrow-Laughlin expression. 
}
\begin{ruledtabular}
\begin{tabular}{ccccc}
RVM & EXD$^{\text{TI}}$ [1] & EXD$^{\text{TI}}$ [2] & EXD$^{\text{TI}}$ [4] & JL \\
\hline
$|1\rangle$ & \underline{0.9294} & -0.3430~ & 0.0903 & 0.8403 \\ 
$|2\rangle$ & -0.1188~ & -0.0693~ & \underline{0.8930} & -0.1086~ \\ 
$|3\rangle$ & 0.0067  &  0.0382 & -0.2596~ & 0.0076 \\ 
$|4\rangle$ & 0.0137  &  0.0191 & -0.0968~ & 0.0395 \\ 
$|5\rangle$ & 0.2540  &  \underline{0.8486} & 0.1519 & 0.4029  \\ 
$|6\rangle$ & 0.0211  &  0.0283 & 0.3097 & 0.0616 \\ 
$|7\rangle$ & -0.2387~ & -0.3935~ & 0.0877 & -0.3380~ \\ 
\end{tabular}
\end{ruledtabular}
\end{table}

\begin{figure}[t]
\centering\includegraphics[width=8.4cm]{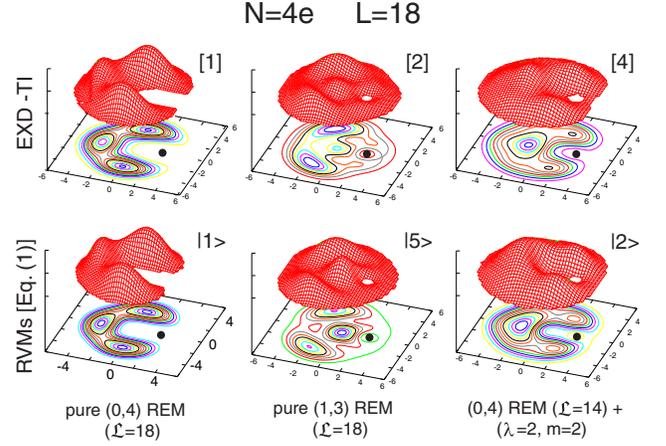}
\caption{(Color online) CPDs for $N=4$ LLL electrons with $L=18$ ($\nu=1/3$). 
Top row: The three lowest-in-energy EXD$^{\text{TI}}$ states (see TABLE \ref{ene_ferm_np4}). 
Bottom row: The RVM trial functions associated with 
the largest expansion coefficients (underlined, see TABLE \ref{exp_coeff}) of 
these three EXD$^{\text{TI}}$ states in the correlated RVM basis. See the text for 
details. The solid dot denotes the fixed point ${\bf r}_0$. Distances in nm. 
}
\label{cpdsexdrvm}
\end{figure}

Of particular interest is the $L=18$ case ($\nu=1/3$) which is considered 
\cite{laug8399} as the  prototype of quantum-liquid states. However, in this case 
we found (see TABLE \ref{ene_ferm_np4}) that the exact 
TI solutions are linear superpositions of the following seven RVM states 
[involving both the (0,4) and (1,3) configurations]:
\begin{eqnarray}
& \hspace{-1cm} 
|1\rangle=\Phi^{\text{REM}}_{18}(0,4), & |2\rangle=\Phi^{\text{REM}}_{14}(0,4) Q_2^2, 
\nonumber \\
& \hspace{-1cm} 
|3\rangle=\Phi^{\text{REM}}_{10}(0,4) Q_2^4, & |4\rangle=\Phi^{\text{REM}}_{6}(0,4) Q_2^6,
\nonumber \\
& \hspace{-1cm} 
|5\rangle=\Phi^{\text{REM}}_{18}(1,3), & |6\rangle=\Phi^{\text{REM}}_{12}(1,3) Q_2^3,
\nonumber \\
& \hspace{-1cm}  
|7\rangle=\Phi^{\text{REM}}_{15}(1,3) Q_3. & ~~~~~
\label{rvmbasisele}
\end{eqnarray}

The expansion coefficients of the three lowest-in-energy EXD $^{\text{TI}}$ states 
(labelled [1], [2], [4]; see TABLE \ref{ene_ferm_np4}) 
in this RVM basis are listed in TABLE \ref{exp_coeff}. One sees that for each 
case, one component (underlined) dominates this expansion; this applies 
for both the yrast state (No. [1]) and the two excitations (Nos. [2] and [4]). 
To further illustrate this, we display in Fig.\ \ref{cpdsexdrvm} the 
conditional probability (pair correlation) distributions (CPDs), 
\begin{equation}
P(z,z_0) = 
\langle \Phi| \sum_{i \neq j}  \delta(z_i -z_0) \delta(z_j-z_0) | \Phi \rangle,
\label{cpds}
\end{equation}
for these three EXD$^{\text{TI}}$ states (top row) and for the 
RVM functions (bottom row) corresponding to
the dominant expansion coefficients. The similarity of the CPDs in each column is 
noticeable and demonstrates that the single RVM functions capture the essence of many-body
correlations in the EXD$^{\text{TI}}$ states. Full quantitative agreement (within 
machine precision) in total energies can be reached by taking into consideration all seven 
RVM basis states [see Eq.\ (\ref{rvmbasisele})]. Naturally, a smaller number of RVM states 
yields intermediate degrees of high-quality quantitative agreement.      

The celebrated Jastrow-Laughlin ansatz \cite{laug8399}, 
\begin{equation}
\Phi^{\text{JL}}[z] = \prod_{1\leq i<j \leq N} (z_i-z_j)^{2p+1}, 
\label{jlwf}
\end{equation}
has been given exclusively an
interpretation of a quantum-fluid state \cite{laug8399,jainbook}. However, since 
the RVM functions span the TI subspace, it follows that any TI trial function
(including the JL ansatz above and the compact CF states) can be expanded
in the RVM basis. As an example, we give in TABLE \ref{exp_coeff} (4th column) 
the RVM expansion of the JL state for $N=4e$ and $L=18$. One sees that, 
compared to the EXD yrast state (1st column), the relative weight of the pure 
(0,4) REM (denoted by $|1\rangle$) is reduced, while the weights of higher-in-energy
vibrational excitations are enhanced. In this context, the liquid 
characteristics are due to the stronger weight of molecular vibrations which 
diminish the rigidity of the molecule.   

\begin{table*}[t] 
\caption{\label{ene_bos_np5}%
LLL spectra of five spinless bosons interacting via a repulsive contact interaction
$g \delta(z_i-z_j)$ . Second column: Dimensions of the full EXD
and the nonspurious TI (in parenthesis) spaces (the EXD space is spanned by
uncorrelated permanents of Darwin-Fock orbitals). Last three columns: Energy
eigenvalues [in units of $g/(\pi \Lambda^2)$] from the diagonalization of the
contact interaction in the TI subspace spanned by the trial functions
$\Phi^{\text{RBM}}_{5k}(0,5) Q_\lambda^m$, $\Phi^{\text{RBM}}_{4k}(1,4) Q_\lambda^m$,
and $\Phi^{\text{RBM}}_{3k}(2,3) Q_\lambda^m$ (RVM digonalization).
Third to sixth columns: the molecular configurations $(n_1,n_2)$ and the
quantum numbers $k$, $\lambda$ and $m$ are indicated within brackets.
There is no nonspurious state with $L=1$. The full EXD spectrum at a given $L$
is constructed by including, in addition to the listed TI energy eigenvalues 
[$D^{\text{TI}}(L)$ in number], all the energies associated with angular momenta smaller
than $L$. An integer in square brackets indicates the energy ordering in the full EXD 
spectrum [including both spurious and TI states (EXD$^{\text{TI}}$)], with [1] denoting 
an yrast state. Eight decimal digits are displayed, but the energy eigenvalues from the 
RVM diagonalization agree with the corresponding EXD$^{\text{TI}}$ ones within machine 
precision.}
\begin{ruledtabular}
\begin{tabular}{llllll}
$L$ & $D^{\text{EXD}}$($D^{\text{TI}}$)& $[(n_1,n_2)\{ k, \lambda, m \}]$ &
\multicolumn{3}{l}{Energy eigenvalues (RVM diag. or EXD$^{\text{TI}}$)} \\
\hline
0 & 1(1) & [(0,5)\{0,$\lambda$,0\}] &
  5.00000000[1]  &   ~~~~      & ~~~~~ \\
2 & 2(1) & [(0,5)\{0,2,1\}] &
  3.75000000[1] &   ~~~~      & ~~~~~~ \\
3 & 3(1) & [(0,5)\{0,3,1\}] &
  3.12500000[1] &   ~~~~      & ~~~~ \\
4 & 5(2) & [(1,4)\{1,$\lambda$,0\}] [(0,5)\{0,2,2\}] &
  2.50000000[1]  &  3.18750000[3] & ~~~~ \\
5 & 7(2) & [(0,5)\{1,$\lambda$,0\}] [(0,5)\{0,5,1\}] & 
  1.87500000[1]  &  3.03125000[3] & ~~~~~ \\
6 & 10(3) &  [(1,4)\{1,2,1\}] [(0,5)\{0,2,3\}] [(0,5)\{0,3,2\}] &
  1.90664171[2] & 2.42925914[3]  & 3.02347415[5] \\
7 & 13(3) & [(1,4)\{1,3,1\}] [(0,5)\{1,2,1\}] [(2,3)\{1,2,2\}] & 
  1.77354877[1]  &  2.07258062[4] &  3.00543311[7] \\
8 & 18(5) & [(1,4)\{2,$\lambda$,0\}] [(1,4)\{1,2,2\}] [(0,5)\{1,3,1\}] &
  1.18821986[1]  & 1.60795253[2]  & 1.91688009[6]  \\
  ~~~ & ~~~~ & [(0,5)\{0,2,4\}] [(2,3)\{2,2,1\}] &
  2.24905854[8] & 3.00273273[11] & ~~~~~~ \\
9 & 23(5) & [(1,4)\{1,5,1\}] [(0,5)\{1,2,2\}] [(2,3)\{3,$\lambda$,0\}]  &
  1.12895814[1] & 1.52195553[3] & 1.89102917[7] \\
  ~~~ & ~~~~ & [(2,3)\{2,3,1\}] [(2,3)\{1,3,2\}] &
  2.06074601[10] & 3.00082677[15] & ~~~~~~ \\
10 & 30(7) & [(0,5)\{2,$\lambda$,0\}] [(0,5)\{1,5,1\}] [(0,5)\{0,5,2\}] &
   0.90026059[1]  &  1.29362646[4]   &  1.51398054[5] \\
~~~~ & ~~~~ & [(1,4)\{2,2,1\}] [(1,4)\{1,2,3\}] [(1,4)\{1,3,2\}]  &
   1.66194766[8]  &  1.95923264[14]  &  2.12274862[17] \\
~~~~ & ~~~~ & [(2,3)\{2,2,2\}]  &
   3.00035191[21]  &   ~~~~      & ~~~~~ \\
11 & 37(7) & [(0,5)\{1,3,2\}] [(0,5)\{1,2,3\}] [(1,4)\{2,3,1\}] &
   1.03755324[2]  &  1.07552423[3]   &  1.50429489[7] \\
~~~~ & ~~~~ & [(2,3)\{3,2,1\}] [(2,3)\{2,5,1\}] [(2,3)\{1,2,4\}]  &
   1.58737738[10]  &  1.94750687[18]  &  2.03365831[20] \\
~~~~ & ~~~~ & [(2,3)\{1,4,2\}]  &
   3.00012024[27]  &   ~~~~      & ~~~~~ \\
12 & 47(10) & [(0,5)\{2,2,1\}] [(1,4)\{3,$\lambda$,0\}] [(1,4)\{2,2,2\}] &
   0.61480761[1]  &  0.93028069[3]   &  1.05066256[5] \\
~~~~ & ~~~~ & [(1,4)\{2,4,1\}] [(1,4)\{1,4,2\}] [(1,4)\{1,2,4\}]  &
   1.34022509[10]  &  1.50000444[11]   &  1.50755634[13] \\
~~~~ & ~~~~ & [(2,3)\{4,$\lambda$,0\}] [(2,3)\{3,3,1\}] [(2,3)\{1,3,3\}] &
   1.64279523[18]  &   1.98164620[27]  &  2.05477689[29]  \\
~~~~ & ~~~~ & [(2,3)\{0,4,3\}]  &
   3.00004768[36]  &   ~~~~~  &  ~~~~~  \\
\end{tabular}
\end{ruledtabular}
\end{table*}

Of great interest also is the $L=30$ ($\nu=1/5$) case, which in the
composite-fermion picture was found to be susceptible to a competition
\cite{chan06} between crystalline and liquid orders. However, we found that the
exact nonspurious states for $L=30$ are actually linear superpositions of the
following 19 $[=D^{\text{TI}}(L=30)]$ RVM functions:
\begin{eqnarray}
\Phi^{\text{REM}}_{6+4k}(0,4)Q_2^{12-2k},\;\; {\text{with}}\;\; k=0,1,2,3,4,5,6;
\nonumber \\
\Phi^{\text{REM}}_{6+3k}(1,3)Q_2^{12-3k/2},\;\; {\text{with}}\;\; k=2,4,6;
\nonumber \\
\Phi^{\text{REM}}_{6+4k}(0,4)Q_3^{8-4k/3},\;\; {\text{with}}\;\; k=0,3; 
\nonumber \\
\Phi^{\text{REM}}_{6+3k}(1,3)Q_3^{8-k},\;\; {\text{with}}\;\; k=2,3,4,5,6,7,8.
\label{l30rvm}
\end{eqnarray}

Diagonalization of the Coulomb interaction in the above TI subspace yielded an
energy 0.25084902 $e^2/(\kappa l_B)$ per electron for the yrast state; this
value agrees again, within machine precision, with the EXD result.
The most sophisticated variants of the composite-fermion theory [including
composite-fermion diagonalization (CFD), composite-fermion crystal (CFC), and
mixed liquid-CFC states \cite{jainbook,chan06,jeon04,jeon07}] fall short in this
respect. Indeed the following higher energies were
found \cite{chan06,note23}: 0.250863(6) (CFD), 0.25094(4) (mixed), 0.25101(4) (CFC).
The CFD basis is not translationally invariant \cite{jeon04,jeon07}.
Consequently, to achieve machine-precision accuracy, the CFD will have to be performed
in the larger space of dimension $D^{\text{EXD}}(L=30)=169$.

\begin{figure}[t]
\centering\includegraphics[width=8.4cm]{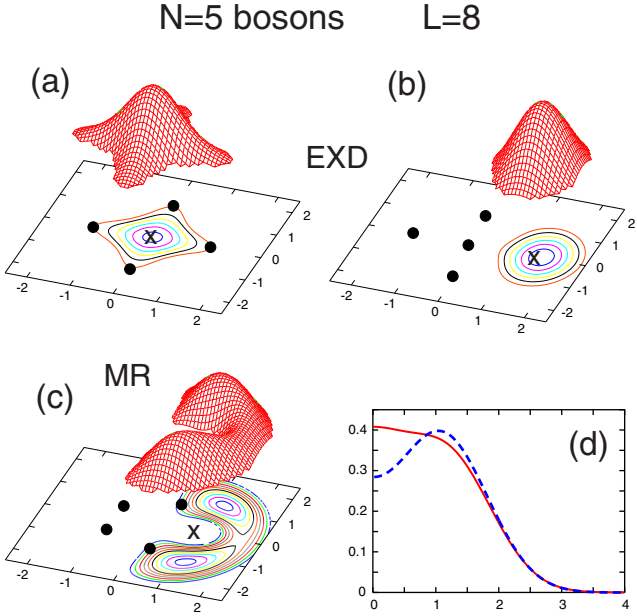}
\caption{(Color online) $N$-body correlation functions for $N=5$ LLL bosons with total 
angular momentum $L=8$. 
(a-b): The yrast EXD$^{\text{TI}}$ state (labeled [1]; see TABLE \ref{ene_bos_np5}). 
(c): The Moore-Read state given by Eq.\ (\ref{mr1}).
(d): Corresponding radial single-particle densities for the yrast EXD$^{\text{TI}}$ 
state (solid line; red online) and the Moore-Read state (dashed line; blue online).
Solid dots denote the four fixed points placed at 
$r_0 \exp[j\pi i/2]$, $j=1,2,3,4$ (with $r_0=1.3 \Lambda$) in (a), $(0,0)$ and
$r_0 \exp[j\pi i/2]$, $j=1,2,3$ (with $r_0=1.3 \Lambda$) in (b), and 
$r_0 \exp[2 j  \pi i/5]$, $j=1,2,3,4$ (with $r_0=1.02 \Lambda$) in (c). 
Crosses denote the expected position of the fifth localized boson according to the 
{\it classical\/} (1,4) polygonal-ring configuration in (a-b) and the (0,5) single-ring
configuration in (c). For the (1,4) configuration, there are two non-equivalent ways of
choosing the four fixed points. In (c), a node, associated with an octupole vibration
(see text), develops at the expected position of the fifth particle. In the radial 
single-particle densities in (d), note the maximum (local minimum) at the origin of the 
solid line (dashed line) in agreement with the underlying (1,4)[(0,5)] configuration.  
} 
\label{nbodyall}
\end{figure}

\subsection{Five bosons}
\label{5bos}

For $N=5$ spinless bosons, one needs to consider rovibrational states [see Eq.\ 
(\ref{mol_trial_wf})] for {\it three\/} distinct molecular configurations, i.e.,
$\Phi^{\text{RBM}}_{5k}(0,5) Q_\lambda^m$, $\Phi^{\text{REM}}_{4k}(1,4) Q_\lambda^m$,
and $\Phi^{\text{REM}}_{3k}(2,3) Q_\lambda^m$. Vibrations with $\lambda \geq 2$ must 
also be considered; since the RVM states are not always orthogonal, a Gram-Schmidt 
orthogonalization is implemented. TABLE \ref{ene_bos_np5} summarizes the quantal 
molecular description in the start of the LLL spectrum ($0 \leq L \leq 12$).

Of particular interest is the $L=8$ case, since it corresponds to the bosonic Moore-Read
state, given \cite{gunn00,coop08} by the analytic expression:
\begin{equation}
\Phi^{\text{MR}}[z]= 
{\cal S} \prod_{i<j \in A} (z_i-z_j)^2 \prod_{k<l \in B} (z_k-z_l)^2,
\label{mr1}
\end{equation}
where sets $A$ and $B$ contain $(N-1)/2$ and $(N+1)/2$ particles, respectively, if $N$ 
is odd. (For $N$ even, both sets contain $N/2$ particles.) ${\cal S}$ symmetrizes over
all possible ways of carrying such a division of $N$ particles into the two sets.

Based on sizeable overlaps with the EXD wave functions \cite{gunn00,coop08}, the 
bosonic Moore-Read states \cite{moor91} are thought to represent the yrast states in the
lowest Landau level with the same angular momentum, i.e., $L=(N-1)^2/2$ for odd $N$
and $L=N(N-2)/2$ for even $N$ \cite{gunn00}.  

However, for $N=5$ and $L=8$, we found (see TABLE \ref{ene_bos_np5}) that the exact
TI solutions are linear superpositions of the following {\it five\/} RVM states
[involving both the (0,5), (1,4), and (2,3) configurations]:
\begin{eqnarray}
& \hspace{-1cm} 
|1\rangle=\Phi^{\text{RBM}}_{8}(1,4),~~~~~~ & |2\rangle=\Phi^{\text{RBM}}_{4}(1,4) Q_2^2,
\nonumber \\
& \hspace{-1cm} 
|3\rangle=\Phi^{\text{RBM}}_{5}(0,5) Q_3,~~ & |4\rangle=\Phi^{\text{RBM}}_{0}(0,5) Q_2^4,
\nonumber \\
& \hspace{-1cm} 
|5\rangle=\Phi^{\text{RBM}}_{6}(2,3) Q_2.~~ & ~~~~~~ 
\label{rvmbasisbos5}
\end{eqnarray}

\begin{table}[b] 
\caption{\label{exp_coeff_n5_bos}%
$N=5$ LLL bosons with $L=8$: Expansion coefficients in the RVM basis 
[Eq.\ (\ref{rvmbasisbos5})] (labelled by the $|i\rangle$'s). The 2nd column gives the yrast 
EXD$^{\text{TI}}$ state (labelled [1]; see TABLE \ref{ene_bos_np5}). The 3rd column 
gives the RVM  expansion coefficients of the corresponding Moore-Read expression
[Eq.\ (\ref{mr1})]. 
}
\begin{ruledtabular}
\begin{tabular}{ccc}
RVM & EXD$^{\text{TI}}$ [1] & MOORE-READ \\
\hline
$|1\rangle$ & \underline{0.7879} & -0.5159~ \\ 
$|2\rangle$ & -0.1162~ &  0.1502  \\ 
$|3\rangle$ & -0.6005~ &  \underline{0.7999} \\ 
$|4\rangle$ & -0.0684~ &  0.1873  \\ 
$|5\rangle$ & -0.0198~ &  -0.1908~ \\ 
\end{tabular}
\end{ruledtabular}
\end{table}

The expansion coefficients of the yrast EXD$^{\text{TI}}$ state (labelled [1]; see TABLE 
\ref{ene_bos_np5}) in this RVM basis are listed in TABLE \ref{exp_coeff_n5_bos}. One 
sees that one component (labelled $|1\rangle$, underlined) dominates this expansion. To further
illustrate this, we display in Fig.\ \ref{nbodyall} the $N$-body correlation 
functions for this EXD$^{\text{TI}}$ state, i.e., the quantity
\begin{equation}
P(z; z_1^0, z_1^0, z_3^0, z_4^0) \propto |\Phi(z; z_1^0, z_2^0, z_3^0, z_4^0)|^2,
\label{nbodyd}
\end{equation}
which gives the probability distribution of finding the fifth boson at a position $z$ 
under the condition that the remaining four bosons are fixed at positions
$z_j^0$, $j=1,2,3,4$. The $N$-body correlations exhibit 
an (1,4) configuration that corresponds to the dominant $\Phi^{\text{RBM}}_{8}(1,4)$
RVM component [component $|1\rangle$, defined in Eq.\ (\ref{rvmbasisbos5})]. 
In particular, when the four fixed points are forming a complete square 
inscribed in a circle of radius $r_0=1.3\Lambda$ [see Fig.\
\ref{nbodyall}(a)], the fifth boson is localized around the origin. When three fixed
points coincide with only three vertices of this square, with the fourth being at the 
origin, the fifth boson is localized around the fourth vertex of the square, as expected 
from the classical (1,4) configuration. 

In contrast to the above, we found that the Moore-Read state [see Eq.\
(\ref{mr1})] exhibits a drastically different behavior in its $N$-body correlation 
function [portrayed in Fig.\ \ref{nbodyall}(c)]. To analyze this behavior, we have 
expanded the Moore-Read state (for $N=5$ bosons and $L=8$) in the RVM basis of 
Eq.\ (\ref{rvmbasisbos5}). As was the case with the Jastrow-Laughlin $1/3$ function for 
$N=4$ electrons, such an expansion is possible due to the fact that the MR state is 
translationally invariant. The corresponding expansion coefficients are listed in TABLE 
\ref{exp_coeff_n5_bos}. It is remarkable that one dominant coefficient (underlined) does 
appear, but it is associated with the RVM basis state $\Phi^{\text{RBM}}_{5}(0,5) Q_3$ 
(denoted as $|3\rangle$), instead of the RVM state $|1\rangle$ that dominates the expansion
of the EXD$^{\text{TI}}$ yrast state. This basis state $|3\rangle$ corresponds to a octupolar
single-phonon vibration of a (0,5) polygonal configuration, and this is reflected in the 
$N$-body correlation plotted in Fig.\ \ref{nbodyall}(c). Indeed, with the four fixed points 
positioned at the vertices of a regular pentagon inscribed in a circle of radius
$r_0=1.02 \Lambda$, the probability of finding the remaining boson is concentrated in the 
neighborhood of the fifth vertex that completes the pentagon (denoted by $X$), but in 
addition it exhibits a prominent node precisely at $X$.

Fig.\ \ref{nbodyall}(d) contrasts the radial single-particle densities of the 
EXD$^{\text{TI}}$ and MR states. Naturally, the radial single-particle densities provide
a reduced amount of information regarding the correlation structure. However,
note that they reflect the underlying (1,4) and (0,5) molecular configurations through the
maximum at the origin of the solid line (EXD$^{\text{TI}}$ state) or the local minimum
at the origin of the dashed line (Moore-Read state), respectively.

The above analysis provides a caveat against drawing conclusions by relying exclusively on
overlaps (as is often the practice in the litterature of fast rotating ultracold bosons
\cite{gunn00,coop08}). Indeed, we can conclude that the Moore-Read state examined here
disagrees in an essential way with the EXD many-body wave functions. 

We stress that the EXD yrast states with the same angular momenta as the MR states
exhibit correlations that conform with the
molecular structures associated with the magic angular momenta defined in Eq.\ (\ref{mam}).
In particular, $L=8$ (corresponding to the MR state for $N=5$) is a magic angular 
momentum associated with a (1,4) configuration (i.e., 8 mod 4 $=$ 0).
Another example is $L=12$ (corresponding to the MR state for $N=6$). In this latter case,
the EXD $N$-body correlation function for the yrast state was studied in Ref.\ 
\cite{baks07}, where it was found (see in paricular Figs.\ 5 and 6 therein) that it 
conformed to a (0,6) polygonal-ring configuration, in agreement with the fact that 
12 mod 6 $=$ 0.

\section{Summary and Discussion}
\label{seccon}

The many-body Hilbert space corresponding to the translationally 
invariant part of the LLL spectra of small systems (whether fermions or bosons,
and for both low and high angular momenta) is spanned by the 
RVM trial functions introduced in Eq.\ (\ref{mol_trial_wf}). 
The yrast and excited states for both short- and long-range interactions can 
always be expressed as linear superpositions of these RVM functions. Thus the 
nature of strong correlations in the LLL reflects the emergence of 
intrinsic point-group symmetries associated with rotations and vibrations of 
molecules formed through particle localization.
We stress the validity of the molecular theory for {\it low\/} angular momenta, 
where "quantum-liquid" physical pictures \cite{laug8399,jainbook,coop08} have been thought to
apply exclusively. Our analysis suggests that liquid-type pictures, associated with 
translationally invariant trial functions (e.g., the Jastrow-Laughlin, compact 
composite-fermion, and Moore-Read functions), are reducible to a description in terms of an 
excited rotating/vibrating quantal molecule. 

\begin{table}[t] 
\caption{\label{cme_rvm_bos_np5_l8}%
Matrix elements of the contact interaction [in units of $g/(\pi \Lambda^2)$] between the RVM
basis states for $N=5$ LLL bosons with $L=8$. The notation for the RVM functions is the same 
as in Eq.\ (\ref{rvmbasisbos5}). Only four decimal points are shown.
}
\begin{ruledtabular}
\begin{tabular}{clllll}
~~~~& \multicolumn{1}{c}{$|1\rangle$} & \multicolumn{1}{c}{$|2\rangle$} 
& \multicolumn{1}{c}{$|3\rangle$} & \multicolumn{1}{c}{$|4\rangle$} 
& \multicolumn{1}{c}{$|5\rangle$}  \\
\hline
$|1\rangle$ 
& ~1.3964  & ~1.5813$\times 10^{-2}$ & ~0.2567 & ~7.7995$\times 10^{-2}$ & ~0.1358 \\
$|2\rangle$ &    & ~2.4933 & -0.1896 & -0.2631 & -0.3711 \\
$|3\rangle$ &    &  & ~1.5554 & ~0.0000 &  ~0.1930 \\
$|4\rangle$ &    &  &  & ~2.4609 & ~0.2492 \\
$|5\rangle$ &    &  &  &  & ~2.0588 \\
\end{tabular}
\end{ruledtabular}
\end{table}

\begin{table*}[t] 
\caption{\label{cme_rvm_bos_np5_l8_en}%
RVM-diagonalization total energies [in units of $g/(\pi \Lambda^2)$], relative errors, and 
overlaps at each intermediate step for the case of the yrast state for $N=5$ LLL bosons with 
$L=8$ (see Sec. \ref{5bos}). The notation for the RVM functions is the same 
as in Eq.\ (\ref{rvmbasisbos5}). The composition of each step is shown in the second column.
The last row displays the corresponding quantities for the Moore-Read function [Eq.\
(\ref{mr1})].
}
\begin{ruledtabular}
\begin{tabular}{ccddd}
Step & Composition & 
\multicolumn{1}{c}{\text{Energy}} & 
\multicolumn{1}{r}{\text{Relative error (\%)}} & 
\multicolumn{1}{r}{\text{Overlap}}  \\
\hline
1 & $|1\rangle$ 
  & 1.3964006 ~\text{(RVM)} & 17.5 & 0.788 \\
2 & $|1\rangle$+$|3\rangle$     
  & 1.2071388 ~\text{(RVM)} & 1.6 & 0.990 \\
3 & $|1\rangle$+$|3\rangle$+$|2\rangle$   
  & 1.1947104 ~\text{(RVM)} & 0.500 & 0.997 \\
4 & $|1\rangle$+$|3\rangle$+$|2\rangle$+$|4\rangle$   
  & 1.1884821 ~\text{(RVM)} & 0.02 & 0.9998 \\
5 & $|1\rangle$+$|3\rangle$+$|2\rangle$+$|4\rangle$+$|5\rangle$
  & 1.1882199 ~\text{(RVM)} & 0.00 & 1.000 \\ 
\multicolumn{2}{c}{MOORE-READ}   
  & 1.2658991  & 6.5 & 0.913  
\end{tabular}
\end{ruledtabular}
\end{table*}

\begin{table*}[t] 
\caption{\label{cme_rvm_rem_np4_l18_en}%
RVM-diagonalization total energies [in units of $e^2/(\kappa l_B)$], relative errors, and 
overlaps at each intermediate step for the case of the yrast state for $N=4$ LLL 
electrons with $L=18$ (see Sec. \ref{4ele}). The notation for the RVM functions is the same 
as in Eq.\ (\ref{rvmbasisele}). The composition of each step is shown in the second column.
The last row displays the corresponding quantities for the Laughlin function [Eq.\ 
(\ref{jlwf})].
}
\begin{ruledtabular}
\begin{tabular}{ccddd}
Step & Composition & 
\multicolumn{1}{c}{\text{Energy}} & 
\multicolumn{1}{r}{\text{Relative error (\%)}} & 
\multicolumn{1}{r}{\text{Overlap}}  \\
\hline
1 & $|1\rangle$ 
  & 1.3217670 ~\text{(RVM)} & 1.23 & 0.929 \\
2 & $|1\rangle$+$|5\rangle$     
  & 1.3174550 ~\text{(RVM)} & 0.90 & 0.961 \\
3 & $|1\rangle$+$|5\rangle$+$|7\rangle$   
  & 1.3081859 ~\text{(RVM)} & 0.19 & 0.992 \\
4 & $|1\rangle$+$|5\rangle$+$|7\rangle$+$|2\rangle$   
  & 1.3059258 ~\text{(RVM)} & 0.015 & 0.99965 \\
5 & $|1\rangle$+$|5\rangle$+$|7\rangle$+$|2\rangle$+$|6\rangle$
  & 1.3058052 ~\text{(RVM)} & 0.006 & 0.99988 \\ 
6 & $|1\rangle$+$|5\rangle$+$|7\rangle$+$|2\rangle$+$|6\rangle+|4\rangle$
  & 1.3057413 ~\text{(RVM)} & 0.001 & 0.99997 \\ 
7 & $|1\rangle$+$|5\rangle$+$|7\rangle$+$|2\rangle$+$|6\rangle+|4\rangle+|3\rangle$
  & 1.3057290 ~\text{(RVM)} & 0.000 & 1.00000 \\ 
\multicolumn{2}{c}{JASTROW-LAUGHLIN}   
  & 1.3105953  & 0.37  & 0.979  
\end{tabular}
\end{ruledtabular}
\end{table*}

In addition to the above conceptual advances, from a computational point of view, 
the introduction of the RVM correlated basis is promising concerning future computational 
developments. Indeed, it has the potential for enabling controlled and
stepwise improvements of the variational wave function, in analogy with previous experiences
from correlated bases in other fields of many-body physics \cite{clark59,clark66,muth00}.
Such developments may enhance computational capabilities for systems with a larger number of
particles than it is currently possible.

The main body of the present paper consisted of two parts. 
The general theoretical background of our 
methodology was presented in Section \ref{secmoltrf}, while Section \ref{secmoldes}
was devoted to case studies. In particular:
 
The analytic trial functions associated with pure rotations of bosonic molecules (i.e., the
RBMs) were derived in Section \ref{secrbm}, followed by a description of the purely
rotational electronic molecular functions (i.e., the REMs) in Section \ref{secrem}. 
Properties of the RBMs and REMs and their association with magic angular momenta were 
discussed in Section \ref{proprbmrem}. The general ro-vibrational trial functions (i.e., the
RVMs) were introduced in Section \ref{secrvm}.

Concerning illustrative examples of the quantal molecular description of the LLL spectra, 
Section \ref{3bos} discussed the case of $N=3$ LLL scalar bosons, while Section \ref{3ele} 
investigated the case of $N=3$ spin-polarized LLL electrons. The case of $N=4$ LLL electrons 
was elaborated in Section \ref{4ele}, along with an analysis of the Jastrow-Laughlin state 
(for fractional filling $\nu=1/3$) from the viewpoint of the present molecular theory. 
Finally, Section \ref{5bos} studied the case of $N=5$ LLL bosons, along with an analysis of 
the Moore-Read state according to the molecular picture. It was shown that the intrinsic
correlation structure of the Moore-Read state disagrees strongly with that of the EXD wave 
function [in spite of having a rather good overlap with the EXD state, calculated by us to 
be 0.913 (see also \cite{gunn00})].

The Appendix discussed in detail the rapid-convergence properties of the RVM basis.

In the case of bosons in harmonic traps, mean-field vortex solutions of the Gross-Pitaevskii
(GP) equation have been proven most useful for the interpretation of experiments in a
variety of situations of Bose-Einstein condensates with very large $N$; see, e.g., Refs.\ 
\cite{dalf99,baym05,fett09}. Consequently, the results presented in this paper, 
and previously \cite{barb06,baks07,yann07}, 
pertaining to crystalline-type correlations in finite systems with a small number $N$ of 
LLL bosons (as well as the earlier prediction of the development of molecular-crystalline 
patterns in non-rotating finite boson systems \cite{roma04}) are unexpected.
While these results, are of intrinsic interest for small
systems, one may also inquire about the size-dependent evolution of the properties of the
system with increasing $N$. This question, and in particular the possibility of a transition 
of LLL bosons to mean-field GP behavior for large $N$, is of high importance conceptually and
practically, and we expect that it will be the focus of future experimental and theoretical 
studies \cite{note7}. 

\begin{acknowledgments}
This work was supported by the U.S. DOE (FG05-86ER45234).
\end{acknowledgments}

\appendix

\section{Convergence in the RVM basis}
\subsection{The example of $N=5$ bosons}


In this section, we analyze in detail the convergence properties of the RVM
diagonalization for the yrast state of $N=5$ bosons with $L=8$ (a case associated with a
Moore-Read function) that was discussed in Sec. \ref{5bos}.

In Table \ref{cme_rvm_bos_np5_l8}, we display the coupling matrix elements for the contact
interaction between the RVM basis functions given in Eq.\ (\ref{rvmbasisbos5}).
The convergence properties in the RVM basis cannot be immediately seen from an inspection
of these coupling matrix elements; while most of the off-diagonal elements in TABLE 
\ref{cme_rvm_bos_np5_l8} are smaller than the differences between the associated diagonal 
elements, a couple of them are indeed larger.

The fast-convergence properties of the RVM basis can be seen through a tabulation of the
intermediate RVM-diagonalization total energies as the size of the RVM basis increases 
in successive steps. In TABLE \ref{cme_rvm_bos_np5_l8_en}, in addition to these intermediate 
RVM energies, we also display the corresponding relative error (relative to the EXD result) 
and the corresponding overlap with the EXD wave function for the yrast state (denoted by the
index [1] in Sec. \ref{5bos}).  
 
We stress that convergence (as a function of the number of the RVM basis functions used in 
the calculation) is seen from TABLE \ref{cme_rvm_bos_np5_l8_en} to be achieved rapidly 
(i.e., already with the use of only two basis functions one obtains a relative error of 
1.6\% for the energy eigenvalue, and a 99\% overlap with the exact eigenfunction). 
In particular, we note that at the second step the RVM wavefunction is already more
accurate compared to the Moore-Read function which exhibits a relative error of 6.5\%
for the energy and an overlap of 91.3\% [see TABLE \ref{cme_rvm_bos_np5_l8_en}].

\subsection{The example of $N=4$ electrons}

In this section, we analyze in detail the convergence properties of the RVM
diagonalization for the yrast state of $N=4$ electrons with $L=18$ (a case associated with a
Jastrow-Laughlin function) that was discussed in Sec. \ref{4ele}.

As previously, the convergence properties of the RVM basis can be seen through a tabulation 
of the intermediate RVM-diagonalization total energies as the size of the RVM basis 
increases in successive steps. In TABLE \ref{cme_rvm_rem_np4_l18_en}, in addition to these 
intermediate RVM energies, we also display the corresponding relative error (relative to the
EXD result) and the corresponding overlap with the EXD wave function for the yrast state 
(denoted by the index [1] in Sec. \ref{4ele}).  
 
Convergence (as a function of the number of the RVM basis functions used in 
the calculation) is seen from TABLE \ref{cme_rvm_rem_np4_l18_en} to be achieved rapidly 
(i.e., already with the use of only two basis functions one obtains a relative error of 
0.90\% for the energy eigenvalue, and a 99\% overlap with the exact eigenfunction). 
In particular, we note that at the third step the RVM wavefunction is already more
accurate compared to the Laughlin function: indeed the latter exhibits a relative error of 
0.37\% for the energy and an overlap of 97.9\%, compared to 0.19\% and 99.2\%, respectively,
in the case of the former [see TABLE \ref{cme_rvm_rem_np4_l18_en}].


\begin{thebibliography}{999}
\bibitem{note3}
An yrast state has the lowest enegy at a given $L$.
\bibitem{note4}
Ref.\ \cite{trug85} multiplied the elementary symmetric polynomials by a 
Pauli-exclusion single-determinant factor to generate a basis spanning the 
fermionic TI subspace, but which has no physical picture associated with it.
\bibitem{note2}
Note that the CF basis and corresponding CF diagonalization have been formulated
only for the case of electrons in the LLL. To date, the application of the CF
methodology to LLL bosons (see. e.g., Refs.\ \cite{coop99,coop08}) is limited to
the simpler mean-field picture (see p. L276 of Ref.\ \cite{jeon04.2}),
which omits residual two-body interactions.
\bibitem{note1}
The lack of translational invariance in the CF basis can be seen from the fact that
for certain pairs, ($L$, $L+1$), of adjacent angular momenta the CF diagonalization yields 
states with the same energy. In this case, the $L+1$ CF state is trivially a 
center-of-mass excitation of the $L$ CF state. For $N=4e$, such pairs of degenerate CF
states are endemic in TABLE I of Ref.\ \cite{jeon07}; for example, $L=10$ and $L=11$
have the same $V_{\text{CF}}$ energy of 1.7849 $e^2/\kappa l_B$, $L=14$ and $L=15$ have 
the same $V_{\text{CF}}$ energy of 1.501 $e^2/\kappa l_B$, etc... Similar ($L$, $L+1$) 
pairs of degenerate CF states are also prominent in the CF yrast energies in all other
cases with $N>4$ presented in Ref.\ \cite{jeon07}. On the contrary, such pairs are
missing from TABLE \ref{ene_ferm_np4} in this paper, which summarizes the energies
according to the RVM diagonalization.
\bibitem{note6}
We note the correspondence $\omega_c \rightarrow 2\Omega$ (see also Ref.\ 
\cite{baks07}).
\bibitem{note82}
In the quantum Hall effect literature, the cusp states are commonly described as 
incompressible, since excitation from each of the corresponding global ground states to the
corresponding lowest excited state entails an energy gap due to the enhanced stability.
\bibitem{note78}
Although translationally invariant, the Hamiltonian in Eq.\ (\ref{hlll}) does not 
separate the center-of-mass coordinate $z_c$ from the internal ones. Such a separation is
always possible if the external potential confinement is harmonic, and this extends also in 
the presence of a magnetic field (or fast rotation) even when the external potential 
confinement can be neglected [case of the Hamiltonian in Eq.\ (\ref{hlll})]. For the case
of $N=3e$ electrons in the LLL, this separation has been carried out explicitly in
Ref.\ \cite{laug83.2} via the introduction of appropriate Jacobi coordinates [see Eqs.\ 
(\ref{jzc}), (\ref{jza}), and (\ref{jzb})]. In this case, only the TI spectrum was involved
(consisting of the eigenenergies of the internal Hamiltonian).
Due to this separability, the general wave-function solution of the Hamiltonian
in Eq.\ (\ref{hlll}) can be written as 
\[ \propto z_c^{L_c} \exp(-z_c^*z_c/2) \Phi^{\text{internal}} [z^J_2,\ldots,z^J_N],\] 
where $z^J_i$, $i=2,\ldots,N$ denote the internal Jacobi coordinates.
Multiplication of this expression by $z_c$ raises the value of the center-of-mass
angular momentum $L_c$ by unity (from $L_c$ to $L_c+1$), without modifying the energy 
of the original state due to the absence of an external confining potential. This accounts
for the spurious degenerate energies (dark solid dots) in Figs.\ \ref{LLLspebos} and 
\ref{LLLspeferm}.
\bibitem{note23}
Numbers shown in parentheses are statistical uncertainties arising from Monte Carlo 
sampling.
\end{thebibliography}
\end{document}